\documentclass[sigconf,balance=false]{acmart}

\usepackage{adjustbox} 
\usepackage{color}     
\usepackage{colortbl}  
\usepackage{hyperref}  
\usepackage{multirow}  
\usepackage{pifont}    
\usepackage{popets}    
     \setcopyright{popets}
     \copyrightyear{YYYY}
     \acmYear{YYYY}
     \acmVolume{YYYY}
     \acmNumber{X}
     \acmDOI{XXXXXXX.XXXXXXX}
     \acmISBN{}
     \acmConference{Proceedings on Privacy Enhancing Technologies}
\usepackage{tabularx}  
\usepackage{todonotes} 
\usepackage{soul}      

\usepackage{glossaries}
\newacronym{api}{API}{Application Programming Interface}
\newacronym{apt}{APT}{Advanced Persistent Threat}
\newacronym{arpanet}{ARPANET}{Advanced Research Projects Agency Network}
\newacronym{ascii}{ASCII}{American Standard Code for Information Interchange}
\newacronym{as}{AS}{Autonomous System}
\newacronym{bgp}{BGP}{Border Gateway Protocol}
\newacronym{bod}{BOD}{Binding Operational Directive}
\newacronym{c2}{C2}{Command and Control}
\newacronym{cc}{C\&C}{Command and Control}
\newacronym{cctld}{ccTLD}{country-code Top Level Domain}
\newacronym{cdf}{CDF}{Cumulative Distribution Function}
\newacronym{cdn}{CDN}{Content Delivery Network}
\newacronym{csk}{CSK}{Combined Signing Key}
\newacronym{dane}{DANE}{DNS-based Authentication of Named Entities}
\newacronym{dbl}{DBL}{Domain Block List}
\newacronym{ddos}{DDoS}{Distributed Denial of Service}
\newacronym{dga}{DGA}{Domain Generation Algorithm}
\newacronym{dhs}{DHS}{Department of Homeland Security}
\newacronym{dkim}{DKIM}{DomainKeys Identified Mail}
\newacronym{dmarc}{DMARC}{Domain-based Message Authentication, Reporting \& Conformance}
\newacronym{dnsbl}{DNSBL}{DNS Block List}
\newacronym{dnssec}{DNSSEC}{DNS Security Extensions}
\newacronym{dns}{DNS}{Domain Name System}
\newacronym{doh}{DoH}{DNS-over-HTTPS}
\newacronym{doq}{DoQ}{DNS-over-QUIC}
\newacronym{dos}{DoS}{Denial of Service}
\newacronym{dot}{DoT}{DNS-over-TLS}
\newacronym{dtls}{DTLS}{Datagram Transport Layer Security}
\newacronym{ec}{EC}{Council of the European Union}
\newacronym{ecs}{ECS}{EDNS(0) Client Subnet}
\newacronym{edns}{EDNS(0)}{Extension Mechanisms for DNS}
\newacronym{eu}{EU}{European Union}
\newacronym{gdpr}{GDPR}{General Data Protection Regulation}
\newacronym{hsts}{HSTS}{HTTP Strict Transport Security}
\newacronym{http}{HTTP}{Hypertext Transfer Protocol}
\newacronym{iana}{IANA}{Internet Assigned Numbers Authority}
\newacronym{icann}{ICANN}{Internet Corporation for Assigned Names and Numbers}
\newacronym{idn}{IDN}{Internationalized Domain Name}
\newacronym{ids}{IDS}{Intrusion Detection System}
\newacronym{ietf}{IETF}{Internet Engineering Task Force}
\newacronym{ip}{IP}{Internet Protocol}
\newacronym{isd}{ISD}{Isolation Domain}
\newacronym{isp}{ISP}{Internet Service Provider}
\newacronym{kisa}{KISA}{Korean Information Security Agency}
\newacronym{ksk}{KSK}{Key Signing Key}
\newacronym{ml}{ML}{Machine Learning}
\newacronym{mta}{MTA}{Mail Transfer Agent}
\newacronym{mtu}{MTU}{Maximum Transmission Unit}
\newacronym{ndn}{NDN}{Named Data Networking}
\newacronym{nren}{NREN}{National Research \& Education Network}
\newacronym{nta}{NTA}{Negative Trust Anchor}
\newacronym{ns}{NS}{Name Server}
\newacronym{ocr}{OCR}{Optical Character Recognition}
\newacronym{octo}{OCTO}{ICANN Office of the CTO}
\newacronym{ofac}{OFAC}{The Office of Foreign Assets Control}
\newacronym{pdns}{pDNS}{passive DNS}
\newacronym{ptr}{PTR}{Pointer Record}
\newacronym{rbl}{RBL}{Real-time Blackhole List}
\newacronym{rddos}{RDDOS}{Reflective Distributed Denial of Service}
\newacronym{rdos}{RDOS}{Reflective Denial of Service}
\newacronym{rfc}{RFC}{Request for Comments}
\newacronym{roa}{ROA}{Route Origin Authorization}
\newacronym{rrl}{RRL}{Response Rate Limiting}
\newacronym{rt}{RT}{Russia Today}
\newacronym{rtbh}{RTBH}{Remote Triggered Black Hole}
\newacronym{rtt}{RTT}{Round-Trip Time}
\newacronym{scp}{SCP}{Subject Certificate Policy}
\newacronym{smtp}{SMTP}{Simple Mail Transfer Protocol}
\newacronym{spf}{SPF}{Sender Policy Framework}
\newacronym{sql}{SQL}{Structured Query Language}
\newacronym{ssac}{SSAC}{Security and Stability Advisory Committe}
\newacronym{ssl}{SSL}{Secure Sockets Layer}
\newacronym{svm}{SVM}{Support Vector Machine}
\newacronym{tld}{TLD}{Top Level Domain}
\newacronym{tls}{TLS}{Transport Layer Security}
\newacronym{ttl}{TTL}{Time To Live}
\newacronym{trr}{TRR}{Trusted Recursive Resolver}
\newacronym{udrp}{UDRP}{Uniform Domain-Name Dispute Resolution Policy}
\newacronym{url}{URL}{Uniform Resource Locator}
\newacronym{usdot}{USDOT}{US Department of Treasury}
\newacronym{ut}{UT}{University of Twente}
\newacronym{vm}{VM}{Virtual Machine}
\newacronym{vp}{VP}{vantage point}
\newacronym{zsk}{ZSK}{Zone Signing Key}

\renewcommand{\todo}[2][]{}

\newcommand{\mmu}[1]{\todo[color=red!40,inline]{moritz: #1}}

\newcolumntype{R}[2]{%
    >{\adjustbox{angle=#1,lap=\width-(#2)}\bgroup}%
    c%
    <{\egroup}%
}
\newcommand*\DomainHead{\multicolumn{1}{R{45}{1em}}}%

\newcommand{\cmark}{\textcolor{green!80!black}{\ding{51}}}
\newcommand{\xmark}{\textcolor{violet}{\ding{55}}}

\definecolor{Gray}{gray}{0.8}

\acmConference{FOCI'24}      
\newcommand\acmPrice[1]{#1}  

\begin{document}

\title[Net Sanctions on RU]{Internet Sanctions on Russian Media: Actions and Effects}


\author{John Kristoff}
\orcid{0000-0002-8994-124X}
\affiliation{%
  \institution{University of Illinois Chicago}
  \city{}
  \country{}
}
\email{jkrist3@uic.edu}

\author{Moritz M\"uller}
\affiliation{
  \institution{SIDN Labs and University of Twente}
  \city{}
  \country{}
}
\email{moritz.muller@sidn.nl}

\author{Arturo Filast\`o}
\affiliation{
  \institution{OONI}
  \city{}
  \country{}
}
\email{arturo@ooni.org}

\author{Max Resing}
\affiliation{
  \institution{University of Twente}
  \city{}
  \country{}
}
\email{m.resing-1@student.utwente.nl}

\author{Chris Kanich}
\affiliation{
  \institution{University of Illinois Chicago}
  \city{}
  \country{}
}
\email{ckanich@uic.edu}

\author{Niels ten Oever}
\affiliation{
  \institution{University of Amsterdam}
  \city{}
  \country{}
}
\email{mail@nielstenoever.net}


\renewcommand{\shortauthors}{Kristoff et al.}
\begin{abstract}

As a response to the Russian aggression against Ukraine, the European
Union (EU), through the notion of `digital sovereignty,' imposed
sanctions on organizations and individuals affiliated with the Russian
Federation that prohibit broadcasting content, including online
distribution. In this paper, we interrogate the implementation of these
sanctions and interpret them as a means to translate the union of states' 
governmental edicts into effective technical countermeasures. Through
longitudinal traffic analysis, we construct an understanding of how ISPs
in different EU countries attempted to enforce these sanctions, and
compare these implementations to similar measures in other western
countries. We find a wide variation of blocking coverage, both
internationally and within individual member states. We draw the
conclusion that digital sovereignty through sanctions in the EU  has a 
concrete but distinctly limited impact on information flows. 

\end{abstract}

\keywords{sanctions, filtering, censorship, Russia}

\maketitle

\section{Introduction}

In response to the Russian aggression against Ukraine, in 2022 the
European Union instated sanctions against ``media outlets under the
permanent direct or indirect control of the leadership of the Russian
Federation'' to ``introduce further restrictive measures to suspend the
broadcasting activities of such media outlets in the Union, or directed
at the Union.'' 

These sanctions are a novel form of government-initiated network
manipulation in several ways: unlike enforcement efforts aimed at e.g.
torrent or streaming sites, the domain names were not seized; unlike
traditional national censorship mechanisms (e.g. China's), the blocking
is being done by a collection of sovereign nations, and is targeted at a
specific, finite set of outlets rather than aiming to be a comprehensive
information control mechanism.

Perhaps most importantly, this effort is not being centrally coordinated
by an individual sovereign entity, but rather by many countries, each
engaging with the internet companies within their own purview. This
event presents an opportunity to investigate a federated, governmental
approach to restricting the flow of internet traffic. We combine several
vantage points and analyses of several different approaches to internet
sanctions to perform a multidimensional characterization of these
actions and their impacts.

This paper makes the following contributions:

\begin{enumerate}
    \item We contribute the first measurement study characterizing
          internet sanctions carried out by a closely coordinating
          collection of states, namely the European Union.
    \item We find that the most widespread sanction
          mechanism is DNS blocking (rather than seizures), and the most
          complete blocking is performed nearest the destination, but that
          blocking itself is far from uniform or ubiquitous, and that
          circumvention via techniques like mirroring is not successfully
          policed.
    \item Synthesizing these results, we conclude that while at a
          governmental level the EU was effectively able to coordinate
          its policy posture with respect to sanctioning these entities,
          the union has not been able to coordinate the technical
          implementation of these sanctions to largely or fully block
          access. While these sanctions no doubt introduced a measurable
          reduction in traffic to the sanctioned entities, complete or
          near-total blockage of sanctioned entities will require new,
          closer forms of coordination at the organizational or technical
          level.
\end{enumerate}

\section{Background}

Transnational communication networks have traditionally been used by
nation states to exert power outside of their territory, while
preventing other nation states from doing so in return
\cite{zajaczReluctantPowerNetworks2019}. To gain control over
information networks, states use different strategies.  Some do so by
engaging in the governance of the internet
\cite{carrPowerPlaysGlobal2015}, such as standard-setting
\cite{ten_oever_making_2022} \cite{denardisGlobalWarInternet2014a} or
policy making around critical internet resources
\cite{christouGainingStakeGlobal2007}
\cite{muellerNetworksStatesGlobal2010}.  However, in these arenas states
need to contend with other actors. In response, several states have made
policy proposals to enhance their 'digital sovereignty'
\cite{coutureWhatDoesNotion2019} or 'data sovereignty'
\cite{muellerWillInternetFragment2017,hummel_data_2021}. Attempts to
limit routes nationally or regionally have thus far largely failed
\cite{donniSchengenRoutingCompliance2015}, but filtering of information
is a commonly used approach \cite{deibertAccessDeniedPractice2008}.

States regularly engage in the unilateral censoring of information on
the internet, and do so in a variety of technical means
\cite{hall_survey_2023}.  Another way of providing instructions for
network operators and infrastructure providers to engage in censorship
is through multilateral internet sanctions.  Some would argue that both
forms of censorship have contributed to permanent internet
fragmentation, which not only complicates technical operations, but
necessitates the need for greater international
cooperation.\cite{cfr_crc_2022}  In their overview paper Drake et al
\cite{drakeInternetFragmentationOverview2016} describe three kinds of
internet fragmentation: commercial, technical, and governmental.
internet sanctions interestingly transverse all these categories.

\subsection{International sanctions}

In the international arena, a sanction is instantiated by a country in
response to the doing of another country. There are different kinds of
sanctions, that range from military actions and sporting events, to
diplomatic and economic sanctions. Here we will focus on economic
sanctions. In economic sanctions a country limits transactions, the
provision of services, or travel by citizens of a particular country, or
particular actors (such as a subsection of the inhabitants of a target
country). In the past sanctions have been placed on telecommunications
equipment as a tool in trade wars
\cite{hongNetworkingChinaDigital2017,rimkevich_economic_2019}. However,
sanctions have not just targeted networking equipment, but also traffic
flows.  An early example of this was documented in a recent report
\cite{badiei_sanctions_2023} that described a case as early as 1999,
when a satellite internet connection provider from the United States
wondered whether it would be in violation of sanctions against
Yugoslavia if it would provide services there. This very early case
clearly stipulates an inherent risk of sanctions, namely an
over-compliance and disproportionate effect on general populations and
therefore their impact on human rights \cite{peksen_better_2009}, which
is problematic because sanctions are regularly invoked in response to
human rights violations \cite{lopez_economic_1997}. To address this
countries often seek to provide carve-outs in sanctions to create more
targeted sanctions that do not negatively impact large populations
\cite{gutmann_precision-guided_2020}. However, these carve-outs do not
always have the desired effect because of over-compliance by the
companies that need to implement these sanctions against particular
actors. Furthermore, companies regularly keep measures they have taken
due to sanctions in place after the sanctions have been lifted, thus
again leading to over-compliance \cite{breen_corporations_2021}.

\subsection{Russia/Ukraine war and EU sanctions}

The current and ongoing aggression against Ukraine started with the
annexation of Crimea and illegal military operations in Ukraine's
eastern Donbas region by the Russian states in February 2014. In
February 2022 Russia started a full scale invasion attempt of Ukraine.

The EU has introduced sanctions against Russia since 2014.  The first
round of EU sanctions were announced in March 2014 and primarily
consisted of travel sanctions. The second round of sanctions in April
2014 were expanded and the EU made it explicit that sanctions were not
aimed at harming people, but designed to bring about change in behavior.
In a third round, more entities and persons were added to the EU
sanctions against Russia which added up to a total of 151 individuals
and 37 entities.  By February 2022, sanctions were applied to Russian
oil and gas, the banking sector, as well as the technology and weapons
industries. These are the heaviest sanctions ever adopted by the EU.

What is most notable from the most recent sanctions is that in March
2022 the EU banned the broadcasting of the news outlets Sputnik and RT.
On June 2 2022, the media outlets Rossiya RTR/RTR Planeta, Rossiya 24
and TV Center International were added as well as the clarification that
Russian state-controlled stations and channels are barred from
distributing their content across the EU, whether via cable, satellite,
internet, or smartphone apps. Furthermore, advertising products or
services on these stations or channels was also forbidden.

\subsection{Research scope}

This paper is exclusively focused on characterizing the impact of the
sanctions passed by the EU on the internet communication of the
sanctioned media entities, its mechanisms, dynamics, and overall
success.  While there have been concrete requests by the government of
Ukraine to internet governance and infrastructure actors ICANN and RIPE,
these fall outside of the remit of this paper. The same is holds true
for initiatives such as the \href{https://wiki.sanctions.net/}{Internet
Sanctions Project} \cite{sanctionsDotNet}, that seeks to provide
guidance for the implementation of internet sanctions for network
operators, and thereby bridging the gap between policy makers and
implementers and limiting over-compliance. This article will also not
focus on the provisioning of numbering and addressing resources to
sanctioned actors, such as those provided by RIPE NCC, the Regional
Internet Registry, which is registered in the Netherlands (and thus
falls under EU law) that covers Europe, the former Soviet Union, and the
Gulf region. 

\subsection{Ethics}

As with many Internet measurement experiments, intentional consideration of ethical ramifications of the work are of the utmost concern. We utilized existing network measurement platforms that ensure active tests are run from systems designed for such purpose and minimize opportunities for abuse.  Measurement platforms that utilized vantage points not under our sole administrative control have informed consent procedures in place and enforce firm restrictions on experiments that may be conducted. However, we raised three unique areas of potential concern not explicitly covered elsewhere.\cite{jones2015ethical,crandall2015forgive}

\begin{enumerate}
  \item Our active measurements may raise sanctions enforcement
        alarms on systems we do not control.
  \item Our active measurements may expose noncompliance with a
        network's sanctions enforcement expectations.
  \item Our active measurement traffic may be unwelcome on the
        infrastructure of a country at war.
\end{enumerate}

Given that our experiments were of modest type, duration, and scope, and that we only attempt to make limited contact to potentially sanctioned resources, we believe that our study  poses no risk on the first concern.  To address concern two we do not highlight any specific networks that may be obligated, but fail to comply with necessary sanctions enforcement requirements.  We also do not publish  the source IP address of the vantage points used in any active measurements.  Regarding the final concern, in addition to carefully construed low-impact active measurement tests, we avoid the use of RIPE Atlas probes and EduVPN exit points located within Ukraine to limit the amount of traffic we place on that country's infrastructure.

Our research study was reviewed by two separate institution review
boards, one in Europe and another in the United States.  They both
concurred with our analysis and approved the experiments.  
\section{Methodology \& Data}\label{sec:methodology}

In this section we provide a high-level overview of our experiments,
measurement methodologies, and collected data.  Our aim is to understand how access to select Russian resources may have been affected due to sanctions enforcement.  We focus on connectivity and access to Russian media organizations from vantage points in Europe unless otherwise noted. We do not examine nor collect any traffic except that which is generated or required by own active measurements through the platforms RIPE Atlas, EduVPN, Dataplane.org and NLNOG RING or provided by OONI.  All active measurements, including those originating from within Ukraine are designed to be low-impact and
nonrecurring. For our RIPE Atlas measurements, we send 12 DNS queries per domain name spread over a period of three hours.
OONI data is retrieved from the public s3 bucket and the raw data is reprocessed using the OONI Data tool\cite{OONIDataKraken}. OONI measurements are collected through their global network of volunteers who have gone through an informed consent procedure where they are informed of the risks associated with participating in this active measurement collection \cite{OONIRisks}.  EduVPN, Dataplane.org, and NLNOG RING measurements are manual non-recurring and each \gls{vp} is accessed sequentially with each measurement run in serial to prevent measurement traffic synchronization toward targets.

\subsection{Sanctioned resource selection}

Block lists of domain names, IP addresses, URLs, or routing information
are commonly used to enforce network operator policies.  In early 2022
we considered two technical proposals that use block list techniques to
enforce internet sanctions against Russia.  One is an ambitious,
community cooperative project focused on transparency.~\cite{sanctionsDotNet}  Another
is a DNS-based firewall approach that blocks access to IP addresses
geo-located to any country under sanction by the US
government.~\cite{sanctionsIp}.  We found no evidence that either
approach has been widely deployed, nor consensus how they should be
deployed.

We then evaluated US and European economic sanction lists published by
national and regional government agencies.  One of the best known and
most influential is from \gls{ofac} in the \gls{usdot}.  \gls{ofac}
maintains and enforces economic sanctions targeting various entities
around the globe, but it is primarily a list of foreign agencies,
commercial organizations, and individuals.~\cite{ofacSanctions} This
list data is populated with names, aliases, and known physical
addresses, but may also include associated internet resources such as as
URLs, email addresses, or cryptocurrency wallet identifiers.  However, we
often found the internet-specific attributes in \gls{ofac} data to be
incomplete, inconsistent, or inaccurate.  Furthermore, we could find no
evidence that the \gls{ofac} list was being widely used for internet
sanctions enforcement.  EU-based sanctions regulations were more scattered
as shown in Table~\ref{tbl:domainnames} in Appendix~\ref{appendix:blockpages}. In
many of these sanctions data sets, we found similar
issues that would make transforming them into internet block list
solutions difficult.

Despite the apparent consistency and specificity challenges with
existing economic sanctions data, multilateral internet sanctions
against Russia began to take shape immediately following the February
24, 2022 attacks on Kyiv.  Implementation details from network providers
were few and far between with some ISPs grudgingly left to work out the
details for themselves.\cite{aa_net_uk_sanctions} We decided to
construct our own list drawn from multiple authoritative sources.  See
Table~\ref{tbl:domainnames}.  The focus on Russian media in our study
reflects the focus of sanctions from official European governing bodies,
but we also include two well-known Russian banks and a branch of the
Russian government that have been sanctioned by the US.  In most of our
experiments we also utilize two \emph{control} web sites that are not
covered by any known sanctions.  One is a static, benign web site on a U.S. academic network.  The other is the \texttt{icanhazip} IP address test
site run by Cloudflare.~\cite{icanhazip}

\subsection{Experiments and measurements}
\label{subsection:experiments_and_measurements}

Internet sanctions enforcement may occur at a variety of points in the
communications path or at different layers in a protocol stack.  To
evaluate enforcement we examine access across four broad dimensions:
reachability, \gls{dns} response, \gls{tls} handshake, and
\gls{http} connection.

\textbf{IP and transport reachability.}
We issue a series of ICMP, TCP, and UDP traceroute probes to our
sanctions list to identify when enforcement occurs at the IP or
transport layers.  Traceroute access failures typically indicate
network-layer enforcement mechanisms such as a packet filter on a
firewall or via a firewalls or black hole route announcement.  Where
applicable, failures above the IP layer by experiments described below
are also recorded.

\textbf{DNS query response behavior.}
For each domain in our sanctions list  we perform both A and AAAA DNS
queries over UDP transport.  Few names have associated AAAA (IPv6)
address mappings and we are limited by each vantage point's local
network configuration whether we can conduct experiments over both IPv4
and IPv6.  Unless otherwise indicated, all results are based on IPv4
transport.  When necessary, we perform identification queries to detect
the resolver configuration if it is not directly available to us.  

We identify block-attempts by relying on fingerprints published 
by OONI~\cite{OONIfingerprints}. Additionally, we manually examine if IP~addresses
point towards websites. If the website contains information about blocking 
efforts, we classify the response as a block-attempt. Finally, we classify 
DNS responses containing errors or non-routable IP addresses (like 127.0.0.1)
as block-attempts.

\textbf{TLS handshake.}
We perform a TLS handshake to the IP addresses associated with port 443
on the targets and perform TLS certificate verification.  We can detect
TLS MiTM attempts by evaluating whether the server X.509 certificate
returned is valid given the SNI and destination IP address in our
requests when validated against the Mozilla root certificate list\cite{MozillaRootCAList}.

\textbf{HTTP\(S\) request.}
Once a TLS session has been established we attempt to retrieve the
content of the homepage by issuing a HTTP GET request for the \texttt{/}
resource.  We issue requests over both HTTP (80) and HTTPS (443) where
applicable.  For each session we record all relevant HTTP response
meta data (e.g., headers and response status code) as well as body
content.

\subsection{Network measurement platforms}

\begin{table}
    \caption{Measurement platforms and supported connectivity tests.}
    \begin{tabularx}{\columnwidth}{l|c|c|c|c}
            \textbf{Platform} & \textbf{IP/TCP} & \textbf{DNS} & \textbf{TLS} & \textbf{HTTP(S)} \\
            \hline
            OONI            & TCP only & \cmark  & \cmark & HTTPS only \\
            \rowcolor{Gray}
            RIPE Atlas      & \xmark   & \cmark   & \xmark & \xmark \\
            EduVPN          & \cmark   & \cmark   & \xmark & \cmark \\
            \rowcolor{Gray}
            Dataplane.org   & \cmark   & \xmark   & \xmark & \cmark \\
            NLNOG RING      & \cmark   & \xmark   & \xmark & \cmark \\
    \end{tabularx}
    \label{tbl:experiments_platforms}
\end{table}

For our study we rely on a variety of network measurement platforms,
summarized in Table~\ref{tbl:experiments_platforms}. Combined, these
platforms allow us to run and evaluate a variety of network experiments
from in-country vantage points. The OONI and RIPE Atlas platforms are
widely used, well understood, and described
elsewhere.\cite{filasto12,staff2015ripe} In OONI most URLs were already part of the testing lists and those missing were added in April 2023\cite{RUMediaPR}. EduVPN, Dataplane.org, and the
NLNOG RING may be less familiar to readers so we briefly summarize them
below.

\textbf{EduVPN}
is a federated VPN project coordinated by SURFnet, the \gls{nren} for the Netherlands.~\cite{EduVPN} Participating organizations can provide two types of access: \emph{Institute Access} and \emph{Secure Internet}.  The former grants access to the internal resources of the host institution.  The latter, which we use, only grants access to the public internet through a trusted server.  Our motivation for using EduVPN is to assess the enforcement of sanctions in academic and research networks.  We setup individual VPN connections at each EduVPN server to appear as a local client on the host network.  DNS resolution configuration varies by institution. We had access to 11 academic networks, of which four were within the EU.

\textbf{Dataplane.org}
is a non-profit network observation and measurement platform that
operates over 300 dedicated and virtual Linux server
systems.~\cite{dataplane}  The majority of vantage points are in hosting provider data
centers around the globe.  While these systems may not reflect end-user
experiences, they help provide additional insight into sanctions
enforcement seen in hosting environments or at the country-level.
Almost all vantage points on this platform utilize Google Public
DNS.\cite{googledns}

\textbf{NLNOG RING}
is project administered by the non-profit Netherlands Network Operator
Group (NLNOG).\cite{nlnogRing} This platform is a collaborative
troubleshooting network consisting of over 600 Linux-based virtual
machines (VMs) in many distinct autonomous systems around the world.
Participating networks contribute VMs and in return are granted access
to all others in the network.  The platform is widely used by network
operators to troubleshoot and debug network-related issues using common
Unix-based tools.  All vantage points use a locally installed DNS
resolver.

\section{Results}

\subsection{A view from OONI Probes}
\label{subsec:resultsooni}
Since we care to know how sanctions enforcement is implemented (via DNS,
TCP/IP, or TLS) we first determine if the DNS responses are consistent.
Given an IP address and domain name pair, we consider an answer to be
DNS consistent if it is possible to successfully establish a TLS
handshake using the domain in the SNI from any vantage
point~\cite{Tsai_2023}.  If we don't do this first, we might
misinterpret the TLS failure as a signal for TLS-level blocking rather
than via the DNS.

For addresses which are not DNS consistent, we manually determine if
they are serving block pages.  We include a sample of block pages in
Appendix \ref{appendix:blockpages}.  If the DNS is consistent, we
proceed to the TCP connection and TLS handshake. For each of these, we
consider a failure to be an indication of blocking and categorize them
based on the specific error condition.

In order to assess the impact of the sanctions from all country origins,
we are interested in understanding how many ISPs in each country are
blocking which sites. Since DNS-based blocking is very prevalent in
Europe\cite{ververis2021understanding}, we introduce an additional
disaggregation based on the configured resolver of the OONI Probes.  We
refer to a resolver configuration as "Internal Resolver", when the
probe's IP address and resolver IP address are both originated from the
same ASN.  Otherwise the resolver is labeled "External Resolver", which
indicates the resolver service is provided by an upstream ISP or
third-party DNS provider.

\begin{figure}[H]
  \centering
  \includegraphics[width=\columnwidth]{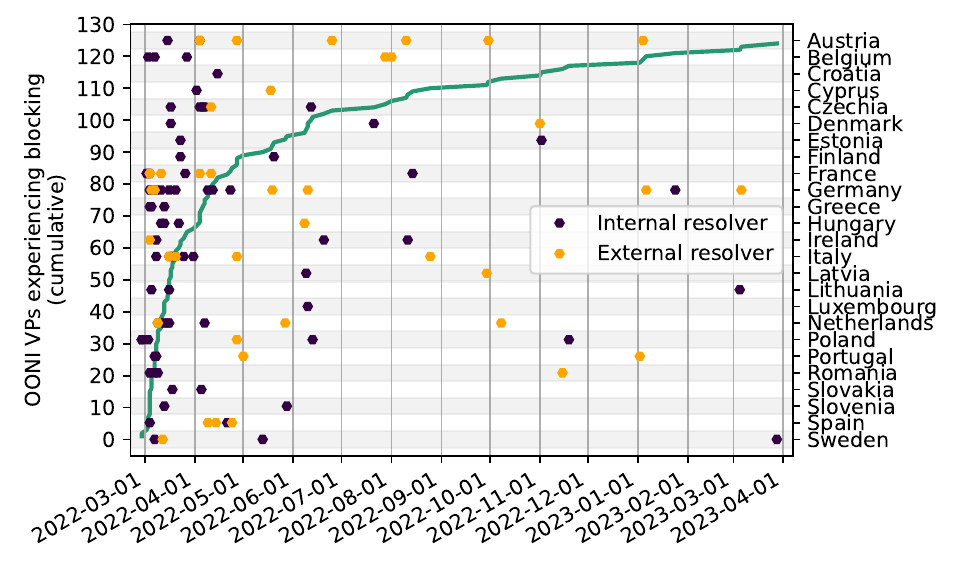}
  \caption{Longitudinal view of first-seen blocking of www.rt.com as observed by OONI. The dark-green is the total number of ASes for which blocking has been observed. 77\% of these ASes enforce sanctions within 3 months.}
  \label{fig:ooni_timeline}
\end{figure} 

In the vast majority of cases an external resolver that performs
blocking is a larger regional upstream ISP.  Since sanctions
enforcement on most public DNS resolver providers is rare, this also allows us to learn how many networks would allow sanctions enforcement to be trivially circumvented by merely switching the resolver to an
alternative service.  We plot the longitudinal results in Figure
\ref{fig:ooni_timeline}.

\begin{figure}[!ht]
  \centering
  \includegraphics[width=\columnwidth]{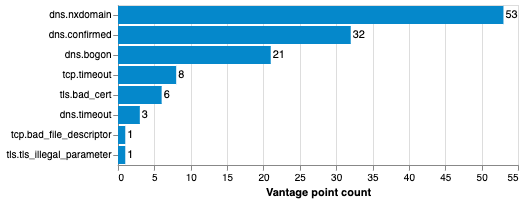}
  \caption{Methods used by ISPs in Europe to implement filtering (OONI Data)}
  \label{fig:ooni_blocking_methods}
\end{figure}

In Figure \ref{fig:ooni_blocking_methods} we illustrate the frequency of
enforcement mechanism types used for \texttt{www.rt.com}, which is the domain
where blocking is most common and for which we have the most
measurements. The error codes are defined in the OONI df-007-errors specification.~\cite{OONISpecDF007} Some additional code are created through some custom analysis. Specifically we mark as \texttt{dns.confirmed} when we see an answer pointing to a known blockpage based on fingerprints in  \cite{OONIBlockingFingerprints}, \texttt{dns.bogon} indicates an answer contains a bogon IP address, while \texttt{tls.bad\_cert} consolidates all the TLS related errors code starting with \texttt{ssl\_}.

\subsection{DNS according to RIPE Atlas}
\label{subsec:resultsripeatlas}
\definecolor{0}{rgb}{0.267004,0.004874,0.329415} 
\definecolor{1}{rgb}{0.282656,0.100196,0.42216} 
\definecolor{2}{rgb}{0.277134,0.185228,0.489898} 
\definecolor{3}{rgb}{0.253935,0.265254,0.529983} 
\definecolor{4}{rgb}{0.221989,0.339161,0.548752} 
\definecolor{5}{rgb}{0.190631,0.407061,0.556089} 
\definecolor{6}{rgb}{0.163625,0.471133,0.558148} 
\definecolor{7}{rgb}{0.139147,0.533812,0.555298} 
\definecolor{8}{rgb}{0.120565,0.596422,0.543611} 
\definecolor{9}{rgb}{0.134692,0.658636,0.517649} 
\definecolor{10}{rgb}{0.20803,0.718701,0.472873} 
\definecolor{11}{rgb}{0.327796,0.77398,0.40664} 
\definecolor{12}{rgb}{0.477504,0.821444,0.318195} 
\definecolor{13}{rgb}{0.647257,0.8584,0.209861} 
\definecolor{14}{rgb}{0.82494,0.88472,0.106217} 
\definecolor{15}{rgb}{0.993248,0.906157,0.143936} 
\renewcommand{\arraystretch}{0.75} 
\begin{table*} 
\setlength\tabcolsep{1pt} 
\caption{Percentage of uncensored DNS responses received by RIPE Atlas probes relying on ISP upstream resolvers.}\label{tbl:country-upstream} 
\centering 
\scriptsize 
\adjustbox{max width=\textwidth}{%
\begin{tabular}{l l c c c c c c c c c c c c c c c c c c c c c c c c c c c} 
 & & \DomainHead{Austria} & \DomainHead{Belgium} & \DomainHead{Bulgaria} & \DomainHead{Croatia} & \DomainHead{Cyprus} & \DomainHead{Czechia} & \DomainHead{Denmark} & \DomainHead{Estonia} & \DomainHead{Finland} & \DomainHead{France} & \DomainHead{Germany} & \DomainHead{Greece} & \DomainHead{Hungary} & \DomainHead{Ireland} & \DomainHead{Italy} & \DomainHead{Lithuania} & \DomainHead{Netherlands} & \DomainHead{Poland} & \DomainHead{Portugal} & \DomainHead{Slovakia} & \DomainHead{Slovenia} & \DomainHead{Spain} & \DomainHead{Sweden} & \DomainHead{United Kingdom} & \DomainHead{Switzerland} & \DomainHead{Russian Federation} & \DomainHead{United States} \\ \midrule 
\multicolumn{2}{l}{\# ASes} &8 &6 &7 &3 &2 &10 &10 &1 &10 &10 &37 &4 &5 &6 &14 &2 &12 &7 &5 &4 &4 &6 &13 &28 &18 &42 &55\\ 
\multicolumn{2}{l}{\# Upstream resolvers} &25 &22 &11 &5 &5 &34 &19 &2 &21 &78 &205 &7 &7 &16 &33 &3 &53 &19 &8 &7 &9 &14 &37 &103 &59 &79 &229\\ 
\multicolumn{2}{l}{\# VPs} &64 &138 &28 &9 &5 &57 &56 &5 &73 &573 &656 &21 &26 &62 &115 &4 &245 &34 &247 &10 &15 &63 &52 &192 &222 &108 &661\\ \midrule 
\parbox[t]{3mm}{\multirow{24}{*}{\rotatebox[origin=c]{90}{Orgs listed by the EC}}} &www.rt.com & \cellcolor{1} \color{white} 7 & \cellcolor{0} \color{white} 1 & \cellcolor{1} \color{white} 9 & \cellcolor{0} \color{white} 0 & \cellcolor{0} \color{white} 0 & \cellcolor{6} \color{white} 38 & \cellcolor{5} \color{white} 33 & \cellcolor{0} \color{white} 0 & \cellcolor{0} \color{white} 2 & \cellcolor{0} \color{white} 2 & \cellcolor{4} \color{white} 25 & \cellcolor{0} \color{white} 0 & \cellcolor{2} \color{white} 14 & \cellcolor{2} \color{white} 13 & \cellcolor{3} \color{white} 23 & \cellcolor{0} \color{white} 0 & \cellcolor{0} \color{white} 5 & \cellcolor{4} \color{white} 28 & \cellcolor{0} \color{white} 0 & \cellcolor{0} \color{white} 0 & \cellcolor{0} \color{white} 0 & \cellcolor{5} \color{white} 34 & \cellcolor{15} 100 & \cellcolor{12} 81 & \cellcolor{15} 99 & \cellcolor{15} 96 & \cellcolor{15} 98 \\ 
 & de.rt.com & \cellcolor{0} \color{white} 6 & \cellcolor{0} \color{white} 1 & \cellcolor{1} \color{white} 9 & \cellcolor{0} \color{white} 0 & \cellcolor{0} \color{white} 0 & \cellcolor{4} \color{white} 30 & \cellcolor{4} \color{white} 31 & \cellcolor{0} \color{white} 0 & \cellcolor{0} \color{white} 2 & \cellcolor{0} \color{white} 4 & \cellcolor{4} \color{white} 25 & \cellcolor{1} \color{white} 9 & \cellcolor{2} \color{white} 14 & \cellcolor{4} \color{white} 28 & \cellcolor{15} 97 & \cellcolor{0} \color{white} 0 & \cellcolor{0} \color{white} 6 & \cellcolor{7} \color{white} 44 & \cellcolor{0} \color{white} 0 &  & \cellcolor{0} \color{white} 0 & \cellcolor{10} 68 & \cellcolor{15} 94 & \cellcolor{12} 81 & \cellcolor{15} 100 & \cellcolor{15} 100 & \cellcolor{15} 98 \\ 
 & deutsch.rt.com & \cellcolor{2} \color{white} 13 & \cellcolor{7} \color{white} 48 & \cellcolor{0} \color{white} 0 & \cellcolor{0} \color{white} 0 & \cellcolor{0} \color{white} 0 & \cellcolor{3} \color{white} 23 & \cellcolor{3} \color{white} 24 & \cellcolor{0} \color{white} 0 & \cellcolor{0} \color{white} 4 & \cellcolor{0} \color{white} 1 & \cellcolor{3} \color{white} 24 & \cellcolor{0} \color{white} 0 & \cellcolor{1} \color{white} 12 & \cellcolor{4} \color{white} 27 & \cellcolor{15} 100 & \cellcolor{0} \color{white} 0 & \cellcolor{9} 62 & \cellcolor{5} \color{white} 35 & \cellcolor{15} 97 & \cellcolor{0} \color{white} 0 & \cellcolor{3} \color{white} 19 & \cellcolor{10} 67 & \cellcolor{15} 100 & \cellcolor{12} 81 & \cellcolor{15} 99 & \cellcolor{15} 96 & \cellcolor{15} 100 \\ 
 & francais.rt.com & \cellcolor{0} \color{white} 4 & \cellcolor{0} \color{white} 3 & \cellcolor{0} \color{white} 0 & \cellcolor{0} \color{white} 0 & \cellcolor{0} \color{white} 0 & \cellcolor{3} \color{white} 21 & \cellcolor{4} \color{white} 25 & \cellcolor{0} \color{white} 0 & \cellcolor{0} \color{white} 2 & \cellcolor{0} \color{white} 3 & \cellcolor{4} \color{white} 26 & \cellcolor{3} \color{white} 22 & \cellcolor{2} \color{white} 14 & \cellcolor{3} \color{white} 22 & \cellcolor{5} \color{white} 34 & \cellcolor{0} \color{white} 0 & \cellcolor{0} \color{white} 6 & \cellcolor{7} \color{white} 46 & \cellcolor{0} \color{white} 0 & \cellcolor{0} \color{white} 0 & \cellcolor{3} \color{white} 19 & \cellcolor{11} 70 & \cellcolor{15} 100 & \cellcolor{12} 80 & \cellcolor{15} 99 & \cellcolor{14} 90 & \cellcolor{15} 98 \\ 
 & fr.rt.com & \cellcolor{1} \color{white} 8 & \cellcolor{7} \color{white} 46 & \cellcolor{1} \color{white} 12 & \cellcolor{0} \color{white} 0 & \cellcolor{0} \color{white} 0 & \cellcolor{4} \color{white} 31 & \cellcolor{4} \color{white} 30 & \cellcolor{0} \color{white} 0 & \cellcolor{0} \color{white} 2 & \cellcolor{0} \color{white} 2 & \cellcolor{14} 93 & \cellcolor{0} \color{white} 0 & \cellcolor{1} \color{white} 11 & \cellcolor{2} \color{white} 14 & \cellcolor{15} 100 & \cellcolor{0} \color{white} 0 & \cellcolor{10} 64 & \cellcolor{5} \color{white} 33 & \cellcolor{15} 96 & \cellcolor{0} \color{white} 0 & \cellcolor{3} \color{white} 22 & \cellcolor{10} 67 & \cellcolor{15} 100 & \cellcolor{13} 83 & \cellcolor{15} 98 & \cellcolor{15} 97 & \cellcolor{15} 99 \\ 
 & actualidad.rt.com & \cellcolor{3} \color{white} 19 & \cellcolor{0} \color{white} 1 & \cellcolor{1} \color{white} 7 & \cellcolor{0} \color{white} 0 & \cellcolor{0} \color{white} 0 & \cellcolor{4} \color{white} 31 & \cellcolor{5} \color{white} 32 & \cellcolor{0} \color{white} 0 & \cellcolor{0} \color{white} 0 & \cellcolor{0} \color{white} 3 & \cellcolor{4} \color{white} 25 & \cellcolor{1} \color{white} 9 & \cellcolor{1} \color{white} 12 & \cellcolor{3} \color{white} 23 & \cellcolor{15} 100 & \cellcolor{0} \color{white} 0 & \cellcolor{0} \color{white} 6 & \cellcolor{6} \color{white} 43 & \cellcolor{0} \color{white} 0 & \cellcolor{0} \color{white} 0 & \cellcolor{0} \color{white} 0 & \cellcolor{10} 66 & \cellcolor{15} 95 & \cellcolor{13} 83 & \cellcolor{15} 100 & \cellcolor{14} 88 & \cellcolor{15} 99 \\ 
 & actualidad-rt.com & \cellcolor{2} \color{white} 14 & \cellcolor{15} 100 & \cellcolor{15} 100 & \cellcolor{15} 100 & \cellcolor{15} 100 & \cellcolor{15} 100 & \cellcolor{15} 100 & \cellcolor{15} 100 & \cellcolor{15} 100 & \cellcolor{15} 100 & \cellcolor{15} 100 & \cellcolor{15} 100 & \cellcolor{15} 100 & \cellcolor{15} 100 & \cellcolor{15} 100 & \cellcolor{15} 100 & \cellcolor{15} 100 & \cellcolor{15} 100 & \cellcolor{15} 100 & \cellcolor{15} 100 & \cellcolor{15} 100 & \cellcolor{15} 97 & \cellcolor{15} 100 & \cellcolor{15} 100 & \cellcolor{15} 100 & \cellcolor{15} 97 & \cellcolor{15} 99 \\ 
 & www.sputniknews.com & \cellcolor{0} \color{white} 4 & \cellcolor{1} \color{white} 8 & \cellcolor{1} \color{white} 9 & \cellcolor{0} \color{white} 0 & \cellcolor{0} \color{white} 0 & \cellcolor{5} \color{white} 33 & \cellcolor{4} \color{white} 26 & \cellcolor{0} \color{white} 0 & \cellcolor{0} \color{white} 6 & \cellcolor{0} \color{white} 0 & \cellcolor{4} \color{white} 30 & \cellcolor{0} \color{white} 0 & \cellcolor{11} 73 & \cellcolor{2} \color{white} 16 & \cellcolor{4} \color{white} 31 & \cellcolor{0} \color{white} 0 & \cellcolor{0} \color{white} 3 & \cellcolor{13} 87 & \cellcolor{0} \color{white} 0 & \cellcolor{0} \color{white} 0 & \cellcolor{0} \color{white} 0 & \cellcolor{15} 100 & \cellcolor{15} 95 & \cellcolor{12} 80 & \cellcolor{15} 100 & \cellcolor{14} 88 & \cellcolor{15} 99 \\ 
 & sputniknewslv.com & \cellcolor{14} 90 & \cellcolor{0} \color{white} 5 & \cellcolor{1} \color{white} 9 & \cellcolor{15} 100 & \cellcolor{0} \color{white} 0 & \cellcolor{4} \color{white} 29 & \cellcolor{4} \color{white} 29 & \cellcolor{0} \color{white} 0 & \cellcolor{0} \color{white} 2 & \cellcolor{9} 60 & \cellcolor{9} 57 & \cellcolor{15} 100 & \cellcolor{7} \color{white} 47 & \cellcolor{3} \color{white} 23 & \cellcolor{4} \color{white} 30 & \cellcolor{15} 100 & \cellcolor{7} \color{white} 49 & \cellcolor{15} 100 & \cellcolor{15} 100 & \cellcolor{0} \color{white} 0 & \cellcolor{3} \color{white} 22 & \cellcolor{15} 100 & \cellcolor{15} 100 & \cellcolor{15} 100 & \cellcolor{15} 100 & \cellcolor{15} 97 & \cellcolor{15} 99 \\ 
 & sputniknews.gr & \cellcolor{15} 100 & \cellcolor{0} \color{white} 1 & \cellcolor{0} \color{white} 0 & \cellcolor{12} 75 & \cellcolor{0} \color{white} 0 & \cellcolor{5} \color{white} 35 & \cellcolor{1} \color{white} 8 & \cellcolor{0} \color{white} 0 & \cellcolor{0} \color{white} 0 & \cellcolor{9} 60 & \cellcolor{10} 63 & \cellcolor{1} \color{white} 11 & \cellcolor{7} \color{white} 46 & \cellcolor{4} \color{white} 25 & \cellcolor{4} \color{white} 26 & \cellcolor{15} 100 & \cellcolor{8} 50 & \cellcolor{15} 100 & \cellcolor{15} 100 & \cellcolor{0} \color{white} 0 & \cellcolor{0} \color{white} 0 & \cellcolor{15} 100 & \cellcolor{15} 100 & \cellcolor{15} 100 & \cellcolor{15} 98 & \cellcolor{15} 96 & \cellcolor{15} 99 \\ 
 & sputniknews.cn & \cellcolor{15} 100 & \cellcolor{0} \color{white} 1 & \cellcolor{1} \color{white} 8 & \cellcolor{12} 80 & \cellcolor{0} \color{white} 0 & \cellcolor{4} \color{white} 27 & \cellcolor{1} \color{white} 8 & \cellcolor{0} \color{white} 0 & \cellcolor{0} \color{white} 2 & \cellcolor{9} 58 & \cellcolor{8} 56 & \cellcolor{15} 100 & \cellcolor{6} \color{white} 41 & \cellcolor{4} \color{white} 25 & \cellcolor{4} \color{white} 31 & \cellcolor{15} 100 & \cellcolor{7} \color{white} 45 & \cellcolor{15} 100 & \cellcolor{15} 100 & \cellcolor{3} \color{white} 19 & \cellcolor{0} \color{white} 0 & \cellcolor{15} 95 & \cellcolor{15} 100 & \cellcolor{15} 100 & \cellcolor{15} 100 & \cellcolor{15} 97 & \cellcolor{15} 100 \\ 
 & radiosputnik.ria.ru & \cellcolor{0} \color{white} 5 & \cellcolor{5} \color{white} 34 & \cellcolor{1} \color{white} 7 & \cellcolor{15} 100 &  & \cellcolor{6} \color{white} 42 & \cellcolor{12} 80 & \cellcolor{0} \color{white} 0 & \cellcolor{0} \color{white} 2 & \cellcolor{15} 99 & \cellcolor{15} 99 & \cellcolor{13} 87 & \cellcolor{15} 100 & \cellcolor{15} 100 & \cellcolor{15} 100 & \cellcolor{0} \color{white} 0 & \cellcolor{15} 100 & \cellcolor{0} \color{white} 6 & \cellcolor{15} 100 & \cellcolor{0} \color{white} 0 & \cellcolor{15} 100 & \cellcolor{15} 95 & \cellcolor{15} 100 & \cellcolor{15} 100 & \cellcolor{15} 100 & \cellcolor{15} 97 & \cellcolor{15} 100 \\ 
 & sputnikglobe.com & \cellcolor{15} 100 & \cellcolor{15} 100 & \cellcolor{15} 100 & \cellcolor{15} 100 & \cellcolor{15} 100 & \cellcolor{15} 100 & \cellcolor{15} 100 & \cellcolor{15} 100 & \cellcolor{1} \color{white} 7 & \cellcolor{15} 100 & \cellcolor{15} 99 & \cellcolor{15} 100 & \cellcolor{15} 100 & \cellcolor{15} 100 & \cellcolor{15} 100 & \cellcolor{15} 100 & \cellcolor{15} 100 & \cellcolor{15} 100 & \cellcolor{15} 100 & \cellcolor{15} 100 & \cellcolor{2} \color{white} 16 & \cellcolor{15} 100 & \cellcolor{15} 100 & \cellcolor{15} 100 & \cellcolor{15} 100 & \cellcolor{14} 89 & \cellcolor{15} 100 \\ 
 & www.rtr-planeta.com & \cellcolor{0} \color{white} 6 & \cellcolor{8} 55 & \cellcolor{15} 100 & \cellcolor{9} 60 & \cellcolor{15} 100 & \cellcolor{15} 100 & \cellcolor{15} 100 & \cellcolor{15} 100 & \cellcolor{0} \color{white} 2 & \cellcolor{15} 95 & \cellcolor{15} 100 & \cellcolor{15} 100 & \cellcolor{15} 100 & \cellcolor{6} \color{white} 41 & \cellcolor{15} 100 & \cellcolor{15} 100 & \cellcolor{7} \color{white} 47 & \cellcolor{1} \color{white} 12 & \cellcolor{0} \color{white} 0 & \cellcolor{15} 100 & \cellcolor{15} 100 & \cellcolor{15} 100 & \cellcolor{15} 100 & \cellcolor{15} 100 & \cellcolor{15} 100 & \cellcolor{15} 97 & \cellcolor{15} 99 \\ 
 & rtr-planeta.ru & \cellcolor{2} \color{white} 17 & \cellcolor{15} 100 & \cellcolor{15} 100 & \cellcolor{15} 100 & \cellcolor{15} 100 & \cellcolor{15} 100 & \cellcolor{12} 77 & \cellcolor{15} 100 & \cellcolor{15} 100 & \cellcolor{15} 100 & \cellcolor{15} 100 & \cellcolor{15} 100 & \cellcolor{15} 100 & \cellcolor{15} 100 & \cellcolor{15} 100 & \cellcolor{15} 100 & \cellcolor{15} 99 & \cellcolor{15} 100 & \cellcolor{15} 100 & \cellcolor{15} 100 & \cellcolor{15} 100 & \cellcolor{15} 100 & \cellcolor{15} 100 & \cellcolor{15} 100 & \cellcolor{15} 100 & \cellcolor{15} 100 & \cellcolor{15} 100 \\ 
 & vgtrk.ru & \cellcolor{15} 100 & \cellcolor{0} \color{white} 3 & \cellcolor{15} 100 & \cellcolor{12} 80 & \cellcolor{15} 100 & \cellcolor{15} 100 & \cellcolor{4} \color{white} 26 & \cellcolor{15} 100 & \cellcolor{0} \color{white} 2 & \cellcolor{15} 100 & \cellcolor{5} \color{white} 33 & \cellcolor{10} 66 & \cellcolor{15} 100 & \cellcolor{15} 100 & \cellcolor{15} 100 & \cellcolor{0} \color{white} 0 & \cellcolor{15} 100 & \cellcolor{5} \color{white} 33 & \cellcolor{15} 96 & \cellcolor{15} 100 & \cellcolor{15} 100 & \cellcolor{15} 100 & \cellcolor{15} 100 & \cellcolor{15} 100 & \cellcolor{15} 100 & \cellcolor{15} 96 & \cellcolor{15} 100 \\ 
 & www.vesti.ru & \cellcolor{2} \color{white} 15 & \cellcolor{8} 52 & \cellcolor{12} 81 & \cellcolor{12} 80 & \cellcolor{15} 100 & \cellcolor{15} 100 & \cellcolor{4} \color{white} 30 & \cellcolor{0} \color{white} 0 & \cellcolor{5} \color{white} 36 & \cellcolor{8} 56 & \cellcolor{15} 100 & \cellcolor{15} 100 & \cellcolor{15} 100 & \cellcolor{5} \color{white} 34 & \cellcolor{15} 100 & \cellcolor{15} 100 & \cellcolor{15} 100 & \cellcolor{6} \color{white} 40 & \cellcolor{15} 96 & \cellcolor{15} 100 & \cellcolor{15} 100 & \cellcolor{15} 100 & \cellcolor{15} 100 & \cellcolor{15} 100 & \cellcolor{15} 100 & \cellcolor{15} 94 & \cellcolor{15} 99 \\ 
 & www.tvc.ru & \cellcolor{3} \color{white} 23 & \cellcolor{0} \color{white} 4 & \cellcolor{12} 81 & \cellcolor{9} 60 & \cellcolor{15} 100 & \cellcolor{15} 100 & \cellcolor{4} \color{white} 28 & \cellcolor{0} \color{white} 0 & \cellcolor{5} \color{white} 35 & \cellcolor{8} 53 & \cellcolor{13} 84 & \cellcolor{5} \color{white} 37 & \cellcolor{15} 100 & \cellcolor{14} 89 & \cellcolor{15} 100 & \cellcolor{0} \color{white} 0 & \cellcolor{7} \color{white} 48 & \cellcolor{15} 100 & \cellcolor{15} 95 & \cellcolor{15} 100 & \cellcolor{15} 100 & \cellcolor{15} 95 & \cellcolor{15} 100 & \cellcolor{15} 100 & \cellcolor{15} 100 & \cellcolor{15} 97 & \cellcolor{15} 100 \\ 
 & ntv.ru & \cellcolor{0} \color{white} 4 & \cellcolor{7} \color{white} 46 & \cellcolor{15} 100 & \cellcolor{0} \color{white} 0 & \cellcolor{15} 100 & \cellcolor{15} 100 & \cellcolor{4} \color{white} 28 & \cellcolor{0} \color{white} 0 & \cellcolor{0} \color{white} 2 & \cellcolor{15} 100 & \cellcolor{4} \color{white} 29 & \cellcolor{12} 75 & \cellcolor{15} 100 & \cellcolor{15} 100 & \cellcolor{15} 100 & \cellcolor{0} \color{white} 0 & \cellcolor{15} 98 & \cellcolor{5} \color{white} 36 & \cellcolor{0} \color{white} 0 & \cellcolor{15} 100 & \cellcolor{15} 100 & \cellcolor{15} 95 & \cellcolor{15} 100 & \cellcolor{15} 98 & \cellcolor{15} 100 & \cellcolor{15} 97 & \cellcolor{15} 100 \\ 
 & smotrim.ru & \cellcolor{15} 100 & \cellcolor{9} 58 & \cellcolor{15} 100 & \cellcolor{3} \color{white} 19 & \cellcolor{15} 100 & \cellcolor{15} 100 & \cellcolor{4} \color{white} 30 & \cellcolor{0} \color{white} 0 & \cellcolor{0} \color{white} 2 & \cellcolor{9} 57 & \cellcolor{4} \color{white} 31 & \cellcolor{8} 50 & \cellcolor{15} 100 & \cellcolor{2} \color{white} 18 & \cellcolor{15} 100 & \cellcolor{0} \color{white} 0 & \cellcolor{15} 100 & \cellcolor{5} \color{white} 33 & \cellcolor{0} \color{white} 0 & \cellcolor{15} 100 & \cellcolor{15} 100 & \cellcolor{15} 95 & \cellcolor{15} 100 & \cellcolor{15} 100 & \cellcolor{15} 100 & \cellcolor{15} 97 & \cellcolor{15} 99 \\ 
 & ren.tv & \cellcolor{1} \color{white} 9 & \cellcolor{0} \color{white} 1 & \cellcolor{15} 100 & \cellcolor{0} \color{white} 0 & \cellcolor{15} 100 & \cellcolor{15} 100 & \cellcolor{5} \color{white} 34 & \cellcolor{0} \color{white} 0 & \cellcolor{0} \color{white} 2 & \cellcolor{15} 99 & \cellcolor{4} \color{white} 29 & \cellcolor{8} 55 & \cellcolor{15} 100 & \cellcolor{5} \color{white} 33 & \cellcolor{15} 100 & \cellcolor{0} \color{white} 0 & \cellcolor{15} 97 & \cellcolor{5} \color{white} 37 & \cellcolor{0} \color{white} 0 & \cellcolor{15} 100 & \cellcolor{15} 100 & \cellcolor{15} 100 & \cellcolor{14} 89 & \cellcolor{15} 100 & \cellcolor{15} 100 & \cellcolor{15} 94 & \cellcolor{15} 99 \\ 
 & 1tv.ru & \cellcolor{0} \color{white} 0 & \cellcolor{0} \color{white} 3 & \cellcolor{15} 100 & \cellcolor{0} \color{white} 0 & \cellcolor{15} 100 & \cellcolor{15} 100 & \cellcolor{4} \color{white} 29 & \cellcolor{0} \color{white} 0 & \cellcolor{0} \color{white} 2 & \cellcolor{15} 99 & \cellcolor{4} \color{white} 31 & \cellcolor{3} \color{white} 19 & \cellcolor{15} 100 & \cellcolor{15} 100 & \cellcolor{15} 100 & \cellcolor{0} \color{white} 0 & \cellcolor{15} 97 & \cellcolor{5} \color{white} 33 & \cellcolor{15} 95 & \cellcolor{12} 80 & \cellcolor{15} 100 & \cellcolor{15} 95 & \cellcolor{15} 100 & \cellcolor{15} 100 & \cellcolor{15} 100 & \cellcolor{15} 100 & \cellcolor{15} 99 \\ 
 & ww.rtarabic.com & \cellcolor{2} \color{white} 15 & \cellcolor{15} 100 & \cellcolor{15} 100 & \cellcolor{15} 100 & \cellcolor{15} 100 & \cellcolor{15} 100 & \cellcolor{7} \color{white} 46 & \cellcolor{0} \color{white} 0 & \cellcolor{6} \color{white} 39 & \cellcolor{8} 56 & \cellcolor{13} 85 & \cellcolor{10} 66 & \cellcolor{15} 100 & \cellcolor{15} 100 & \cellcolor{15} 100 & \cellcolor{15} 100 & \cellcolor{15} 96 & \cellcolor{15} 100 & \cellcolor{15} 100 & \cellcolor{15} 100 & \cellcolor{15} 100 & \cellcolor{15} 100 & \cellcolor{15} 100 & \cellcolor{15} 100 & \cellcolor{15} 100 & \cellcolor{15} 96 & \cellcolor{15} 100 \\ 
 & sputnikarabic.ae & \cellcolor{3} \color{white} 19 & \cellcolor{15} 100 & \cellcolor{15} 100 & \cellcolor{15} 100 & \cellcolor{15} 100 & \cellcolor{15} 100 & \cellcolor{4} \color{white} 25 & \cellcolor{0} \color{white} 0 & \cellcolor{0} \color{white} 2 & \cellcolor{9} 58 & \cellcolor{7} \color{white} 47 & \cellcolor{8} 50 & \cellcolor{15} 100 & \cellcolor{15} 100 & \cellcolor{15} 100 & \cellcolor{15} 100 & \cellcolor{7} \color{white} 45 & \cellcolor{15} 100 & \cellcolor{15} 95 & \cellcolor{15} 100 & \cellcolor{15} 100 & \cellcolor{15} 100 & \cellcolor{15} 100 & \cellcolor{15} 100 & \cellcolor{15} 100 & \cellcolor{15} 96 & \cellcolor{15} 99 \\ 
\midrule 
\parbox[t]{3mm}{\multirow{11}{*}{\rotatebox[origin=c]{90}{Mirror pages}}} &esrt.online & \cellcolor{2} \color{white} 17 & \cellcolor{15} 100 & \cellcolor{15} 100 & \cellcolor{15} 100 & \cellcolor{15} 100 & \cellcolor{15} 94 & \cellcolor{15} 100 & \cellcolor{15} 100 & \cellcolor{15} 100 & \cellcolor{15} 99 & \cellcolor{15} 99 & \cellcolor{15} 100 & \cellcolor{14} 92 & \cellcolor{15} 100 & \cellcolor{15} 100 & \cellcolor{15} 100 & \cellcolor{15} 100 & \cellcolor{15} 100 & \cellcolor{15} 100 & \cellcolor{15} 100 & \cellcolor{15} 100 & \cellcolor{15} 100 & \cellcolor{15} 100 & \cellcolor{15} 100 & \cellcolor{15} 100 & \cellcolor{15} 94 & \cellcolor{15} 99 \\ 
 & esrt.press & \cellcolor{4} \color{white} 26 & \cellcolor{15} 100 & \cellcolor{15} 100 & \cellcolor{15} 100 &  & \cellcolor{15} 100 & \cellcolor{15} 100 & \cellcolor{15} 100 & \cellcolor{15} 100 & \cellcolor{15} 99 & \cellcolor{15} 100 & \cellcolor{15} 100 & \cellcolor{15} 100 & \cellcolor{15} 100 & \cellcolor{15} 100 & \cellcolor{15} 100 & \cellcolor{15} 100 & \cellcolor{15} 100 & \cellcolor{15} 100 & \cellcolor{15} 100 & \cellcolor{15} 100 & \cellcolor{15} 100 & \cellcolor{15} 100 & \cellcolor{15} 100 & \cellcolor{15} 100 & \cellcolor{15} 100 & \cellcolor{15} 100 \\ 
 & rtde.site & \cellcolor{2} \color{white} 14 & \cellcolor{15} 100 & \cellcolor{15} 100 & \cellcolor{15} 100 & \cellcolor{15} 100 & \cellcolor{15} 100 & \cellcolor{12} 76 & \cellcolor{15} 100 & \cellcolor{15} 100 & \cellcolor{15} 99 & \cellcolor{4} \color{white} 28 & \cellcolor{12} 75 & \cellcolor{15} 100 & \cellcolor{15} 100 & \cellcolor{15} 100 & \cellcolor{15} 100 & \cellcolor{15} 100 & \cellcolor{15} 100 & \cellcolor{15} 96 & \cellcolor{15} 100 & \cellcolor{15} 100 & \cellcolor{15} 100 & \cellcolor{15} 100 & \cellcolor{15} 100 & \cellcolor{15} 99 & \cellcolor{15} 100 & \cellcolor{15} 100 \\ 
 & rtde.xyz & \cellcolor{0} \color{white} 0 & \cellcolor{15} 100 & \cellcolor{15} 100 & \cellcolor{15} 100 & \cellcolor{15} 100 & \cellcolor{15} 100 & \cellcolor{11} 73 & \cellcolor{15} 100 & \cellcolor{15} 100 & \cellcolor{15} 99 & \cellcolor{4} \color{white} 30 & \cellcolor{8} 55 & \cellcolor{15} 100 & \cellcolor{15} 100 & \cellcolor{15} 100 & \cellcolor{15} 100 & \cellcolor{15} 100 & \cellcolor{15} 100 & \cellcolor{15} 98 & \cellcolor{15} 100 & \cellcolor{15} 100 & \cellcolor{15} 100 & \cellcolor{15} 100 & \cellcolor{15} 100 & \cellcolor{15} 100 & \cellcolor{15} 94 & \cellcolor{15} 99 \\ 
 & rtde.team & \cellcolor{0} \color{white} 0 & \cellcolor{15} 100 & \cellcolor{15} 100 & \cellcolor{15} 100 & \cellcolor{15} 100 & \cellcolor{15} 100 & \cellcolor{11} 73 & \cellcolor{15} 100 & \cellcolor{15} 100 & \cellcolor{15} 100 & \cellcolor{5} \color{white} 32 & \cellcolor{8} 50 & \cellcolor{15} 100 & \cellcolor{15} 100 & \cellcolor{15} 100 & \cellcolor{15} 100 & \cellcolor{15} 100 & \cellcolor{15} 100 & \cellcolor{15} 98 & \cellcolor{15} 100 & \cellcolor{15} 100 & \cellcolor{15} 100 & \cellcolor{15} 100 & \cellcolor{15} 100 & \cellcolor{15} 100 & \cellcolor{15} 96 & \cellcolor{15} 99 \\ 
 & test.rtde.live & \cellcolor{3} \color{white} 22 & \cellcolor{15} 100 & \cellcolor{15} 100 & \cellcolor{15} 100 & \cellcolor{15} 100 & \cellcolor{15} 100 & \cellcolor{12} 76 & \cellcolor{15} 100 & \cellcolor{15} 100 & \cellcolor{15} 100 & \cellcolor{4} \color{white} 25 & \cellcolor{8} 54 & \cellcolor{15} 100 & \cellcolor{15} 100 & \cellcolor{15} 100 & \cellcolor{15} 100 & \cellcolor{15} 100 & \cellcolor{15} 100 & \cellcolor{15} 96 & \cellcolor{15} 100 & \cellcolor{15} 100 & \cellcolor{15} 94 & \cellcolor{15} 100 & \cellcolor{15} 100 & \cellcolor{15} 100 & \cellcolor{15} 100 & \cellcolor{15} 99 \\ 
 & rtde.live & \cellcolor{2} \color{white} 18 & \cellcolor{15} 98 & \cellcolor{15} 100 & \cellcolor{15} 100 & \cellcolor{15} 100 & \cellcolor{15} 100 & \cellcolor{12} 76 & \cellcolor{15} 100 & \cellcolor{15} 100 & \cellcolor{15} 99 & \cellcolor{15} 98 & \cellcolor{14} 92 & \cellcolor{15} 100 & \cellcolor{15} 100 & \cellcolor{15} 100 & \cellcolor{15} 100 & \cellcolor{15} 100 & \cellcolor{15} 100 & \cellcolor{15} 100 & \cellcolor{15} 100 & \cellcolor{15} 100 & \cellcolor{15} 95 & \cellcolor{15} 100 & \cellcolor{15} 100 & \cellcolor{14} 93 & \cellcolor{15} 97 & \cellcolor{15} 100 \\ 
 & test.rtde.website & \cellcolor{15} 100 & \cellcolor{15} 100 & \cellcolor{15} 100 & \cellcolor{15} 100 & \cellcolor{15} 100 & \cellcolor{15} 100 & \cellcolor{12} 81 & \cellcolor{15} 100 & \cellcolor{15} 100 & \cellcolor{15} 100 & \cellcolor{3} \color{white} 24 & \cellcolor{9} 60 & \cellcolor{15} 100 & \cellcolor{15} 100 & \cellcolor{15} 100 & \cellcolor{15} 100 & \cellcolor{15} 100 & \cellcolor{15} 100 & \cellcolor{15} 96 & \cellcolor{15} 100 & \cellcolor{15} 100 & \cellcolor{15} 100 & \cellcolor{15} 100 & \cellcolor{15} 100 & \cellcolor{15} 100 & \cellcolor{15} 100 & \cellcolor{15} 100 \\ 
 & rtde.tech & \cellcolor{1} \color{white} 12 & \cellcolor{15} 100 & \cellcolor{15} 100 & \cellcolor{15} 100 & \cellcolor{15} 100 & \cellcolor{15} 100 & \cellcolor{13} 85 & \cellcolor{15} 100 & \cellcolor{15} 100 & \cellcolor{15} 100 & \cellcolor{4} \color{white} 27 & \cellcolor{11} 72 & \cellcolor{15} 100 & \cellcolor{15} 100 & \cellcolor{15} 100 & \cellcolor{15} 100 & \cellcolor{15} 100 & \cellcolor{15} 100 & \cellcolor{15} 96 & \cellcolor{15} 100 & \cellcolor{15} 100 & \cellcolor{15} 100 & \cellcolor{15} 100 & \cellcolor{15} 100 & \cellcolor{15} 100 & \cellcolor{15} 97 & \cellcolor{15} 99 \\ 
 & rtde.world & \cellcolor{5} \color{white} 35 & \cellcolor{15} 100 & \cellcolor{15} 100 & \cellcolor{15} 100 & \cellcolor{15} 100 & \cellcolor{15} 100 & \cellcolor{12} 78 & \cellcolor{15} 100 & \cellcolor{15} 100 & \cellcolor{15} 99 & \cellcolor{4} \color{white} 29 & \cellcolor{10} 63 & \cellcolor{15} 100 & \cellcolor{15} 100 & \cellcolor{15} 100 & \cellcolor{15} 100 & \cellcolor{15} 100 & \cellcolor{15} 100 & \cellcolor{15} 94 &  & \cellcolor{15} 100 & \cellcolor{15} 100 & \cellcolor{15} 100 & \cellcolor{15} 100 & \cellcolor{15} 100 & \cellcolor{15} 100 & \cellcolor{15} 99 \\ 
 & rtde.me & \cellcolor{3} \color{white} 21 & \cellcolor{15} 100 & \cellcolor{15} 100 & \cellcolor{15} 100 & \cellcolor{15} 100 & \cellcolor{15} 100 & \cellcolor{12} 76 & \cellcolor{15} 100 & \cellcolor{15} 100 & \cellcolor{15} 99 & \cellcolor{4} \color{white} 29 & \cellcolor{7} \color{white} 46 & \cellcolor{15} 100 & \cellcolor{15} 100 & \cellcolor{15} 100 & \cellcolor{15} 100 & \cellcolor{15} 100 & \cellcolor{15} 100 & \cellcolor{15} 95 & \cellcolor{15} 100 & \cellcolor{15} 100 & \cellcolor{15} 100 & \cellcolor{15} 100 & \cellcolor{15} 100 & \cellcolor{15} 98 & \cellcolor{15} 97 & \cellcolor{15} 99 \\ 
\midrule 
\parbox[t]{3mm}{\multirow{9}{*}{\rotatebox[origin=c]{90}{TV streaming svcs}}} &a-russia.ru & \cellcolor{15} 100 & \cellcolor{15} 100 & \cellcolor{15} 100 & \cellcolor{9} 60 & \cellcolor{15} 100 & \cellcolor{15} 94 & \cellcolor{13} 86 & \cellcolor{0} \color{white} 0 & \cellcolor{15} 100 & \cellcolor{15} 100 & \cellcolor{4} \color{white} 30 & \cellcolor{8} 50 & \cellcolor{15} 100 & \cellcolor{15} 100 & \cellcolor{15} 100 & \cellcolor{0} \color{white} 0 & \cellcolor{15} 100 & \cellcolor{15} 100 & \cellcolor{15} 95 & \cellcolor{15} 100 & \cellcolor{15} 100 & \cellcolor{15} 100 & \cellcolor{15} 100 & \cellcolor{15} 99 & \cellcolor{15} 100 & \cellcolor{15} 100 & \cellcolor{15} 98 \\ 
 & wwitv.com & \cellcolor{15} 100 & \cellcolor{15} 100 & \cellcolor{15} 100 & \cellcolor{15} 100 & \cellcolor{15} 100 & \cellcolor{15} 100 & \cellcolor{6} \color{white} 43 & \cellcolor{15} 100 & \cellcolor{14} 88 & \cellcolor{15} 100 & \cellcolor{4} \color{white} 28 & \cellcolor{4} \color{white} 28 & \cellcolor{15} 100 & \cellcolor{15} 100 & \cellcolor{15} 100 & \cellcolor{0} \color{white} 0 & \cellcolor{15} 100 & \cellcolor{15} 100 & \cellcolor{15} 94 & \cellcolor{15} 100 & \cellcolor{15} 100 & \cellcolor{15} 100 & \cellcolor{15} 100 & \cellcolor{15} 100 & \cellcolor{15} 100 & \cellcolor{15} 100 & \cellcolor{15} 100 \\ 
 & www.glaz.tv & \cellcolor{15} 100 & \cellcolor{15} 100 & \cellcolor{15} 100 & \cellcolor{9} 60 & \cellcolor{15} 100 & \cellcolor{15} 100 & \cellcolor{12} 81 & \cellcolor{0} \color{white} 0 & \cellcolor{15} 97 & \cellcolor{15} 100 & \cellcolor{6} \color{white} 43 & \cellcolor{9} 60 & \cellcolor{15} 100 & \cellcolor{15} 100 & \cellcolor{15} 100 & \cellcolor{0} \color{white} 0 & \cellcolor{15} 100 & \cellcolor{15} 100 & \cellcolor{15} 96 & \cellcolor{15} 100 & \cellcolor{15} 100 & \cellcolor{15} 100 & \cellcolor{13} 83 & \cellcolor{15} 99 & \cellcolor{15} 100 & \cellcolor{15} 100 & \cellcolor{15} 100 \\ 
 & www.russisches-tv-fernsehen.de & \cellcolor{15} 100 & \cellcolor{15} 100 & \cellcolor{15} 100 & \cellcolor{9} 60 & \cellcolor{15} 100 & \cellcolor{15} 100 & \cellcolor{15} 100 & \cellcolor{15} 100 & \cellcolor{15} 100 & \cellcolor{15} 100 & \cellcolor{15} 100 & \cellcolor{15} 100 & \cellcolor{15} 100 & \cellcolor{15} 100 & \cellcolor{15} 100 & \cellcolor{0} \color{white} 0 & \cellcolor{15} 100 & \cellcolor{15} 100 & \cellcolor{15} 100 & \cellcolor{15} 100 & \cellcolor{15} 100 & \cellcolor{15} 100 & \cellcolor{15} 100 & \cellcolor{15} 100 & \cellcolor{15} 100 & \cellcolor{15} 96 & \cellcolor{15} 99 \\ 
 & ontvtime.tv & \cellcolor{15} 100 & \cellcolor{8} 53 & \cellcolor{15} 100 & \cellcolor{9} 60 & \cellcolor{15} 100 & \cellcolor{15} 100 & \cellcolor{4} \color{white} 31 & \cellcolor{0} \color{white} 0 & \cellcolor{14} 88 & \cellcolor{15} 100 & \cellcolor{4} \color{white} 31 & \cellcolor{11} 71 & \cellcolor{15} 100 & \cellcolor{4} \color{white} 25 & \cellcolor{15} 100 & \cellcolor{0} \color{white} 0 & \cellcolor{15} 100 & \cellcolor{15} 100 & \cellcolor{15} 95 & \cellcolor{15} 100 & \cellcolor{15} 100 & \cellcolor{15} 100 & \cellcolor{15} 100 & \cellcolor{15} 100 & \cellcolor{15} 99 & \cellcolor{15} 100 & \cellcolor{15} 100 \\ 
 & spbtv.online & \cellcolor{15} 100 & \cellcolor{15} 100 & \cellcolor{15} 100 & \cellcolor{15} 100 & \cellcolor{15} 100 & \cellcolor{15} 100 & \cellcolor{15} 100 & \cellcolor{0} \color{white} 0 & \cellcolor{15} 100 & \cellcolor{15} 100 & \cellcolor{5} \color{white} 32 & \cellcolor{8} 50 & \cellcolor{15} 100 & \cellcolor{15} 100 & \cellcolor{15} 100 & \cellcolor{0} \color{white} 0 & \cellcolor{15} 100 & \cellcolor{15} 100 & \cellcolor{15} 97 & \cellcolor{15} 100 & \cellcolor{15} 100 & \cellcolor{15} 100 & \cellcolor{15} 100 & \cellcolor{15} 100 & \cellcolor{15} 100 & \cellcolor{15} 97 & \cellcolor{15} 100 \\ 
 & www.coolstreaming.us & \cellcolor{15} 100 & \cellcolor{15} 100 & \cellcolor{15} 100 & \cellcolor{15} 100 & \cellcolor{15} 100 & \cellcolor{15} 100 & \cellcolor{15} 100 & \cellcolor{15} 100 & \cellcolor{15} 100 & \cellcolor{15} 100 & \cellcolor{15} 100 & \cellcolor{15} 100 & \cellcolor{15} 100 & \cellcolor{15} 100 & \cellcolor{15} 100 & \cellcolor{15} 100 & \cellcolor{15} 100 & \cellcolor{15} 100 & \cellcolor{15} 100 & \cellcolor{15} 100 & \cellcolor{15} 100 & \cellcolor{15} 100 & \cellcolor{14} 92 & \cellcolor{15} 99 & \cellcolor{15} 100 & \cellcolor{15} 97 & \cellcolor{15} 99 \\ 
 & www.livehdtv.net & \cellcolor{15} 94 & \cellcolor{15} 100 & \cellcolor{15} 100 & \cellcolor{15} 100 & \cellcolor{15} 100 & \cellcolor{15} 100 & \cellcolor{15} 100 & \cellcolor{0} \color{white} 0 & \cellcolor{15} 100 & \cellcolor{15} 99 & \cellcolor{6} \color{white} 43 & \cellcolor{5} \color{white} 37 & \cellcolor{15} 100 & \cellcolor{15} 100 & \cellcolor{15} 100 & \cellcolor{0} \color{white} 0 & \cellcolor{15} 100 & \cellcolor{15} 100 & \cellcolor{15} 96 & \cellcolor{15} 100 & \cellcolor{15} 100 & \cellcolor{15} 100 & \cellcolor{15} 100 & \cellcolor{15} 100 & \cellcolor{15} 100 & \cellcolor{15} 96 & \cellcolor{15} 99 \\ 
 & snanews.de & \cellcolor{2} \color{white} 15 & \cellcolor{0} \color{white} 1 & \cellcolor{1} \color{white} 9 & \cellcolor{15} 100 & \cellcolor{4} \color{white} 25 & \cellcolor{8} 50 & \cellcolor{4} \color{white} 28 & \cellcolor{15} 100 & \cellcolor{0} \color{white} 2 & \cellcolor{9} 59 & \cellcolor{4} \color{white} 31 & \cellcolor{4} \color{white} 28 & \cellcolor{2} \color{white} 15 & \cellcolor{4} \color{white} 26 & \cellcolor{13} 83 & \cellcolor{15} 100 & \cellcolor{13} 86 & \cellcolor{15} 100 & \cellcolor{15} 94 & \cellcolor{0} \color{white} 0 & \cellcolor{4} \color{white} 30 & \cellcolor{15} 100 & \cellcolor{15} 100 & \cellcolor{15} 100 & \cellcolor{15} 100 & \cellcolor{15} 96 & \cellcolor{15} 100 \\ 
\midrule 
\parbox[t]{3mm}{\multirow{4}{*}{\rotatebox[origin=c]{90}{Other}}} &duma.gov.ru & \cellcolor{15} 100 & \cellcolor{15} 100 & \cellcolor{15} 100 & \cellcolor{15} 100 & \cellcolor{15} 100 & \cellcolor{15} 100 & \cellcolor{12} 81 & \cellcolor{15} 100 & \cellcolor{15} 100 & \cellcolor{15} 100 & \cellcolor{15} 100 & \cellcolor{15} 100 & \cellcolor{15} 100 & \cellcolor{15} 100 & \cellcolor{15} 100 & \cellcolor{15} 100 & \cellcolor{15} 100 & \cellcolor{15} 100 & \cellcolor{15} 100 & \cellcolor{15} 100 & \cellcolor{15} 100 & \cellcolor{15} 100 & \cellcolor{15} 100 & \cellcolor{15} 100 & \cellcolor{15} 100 & \cellcolor{15} 97 & \cellcolor{15} 99 \\ 
 & www.sber-bank.by & \cellcolor{15} 100 & \cellcolor{15} 100 & \cellcolor{15} 100 & \cellcolor{15} 100 & \cellcolor{15} 100 & \cellcolor{15} 100 & \cellcolor{12} 77 & \cellcolor{15} 100 & \cellcolor{15} 100 & \cellcolor{15} 100 & \cellcolor{15} 100 & \cellcolor{15} 100 & \cellcolor{15} 100 & \cellcolor{15} 100 & \cellcolor{15} 100 & \cellcolor{15} 100 & \cellcolor{15} 100 & \cellcolor{15} 100 & \cellcolor{15} 100 & \cellcolor{15} 100 & \cellcolor{15} 100 & \cellcolor{15} 100 & \cellcolor{15} 100 & \cellcolor{15} 100 & \cellcolor{15} 100 & \cellcolor{15} 94 & \cellcolor{15} 100 \\ 
 & www.sberbank.ru & \cellcolor{15} 100 & \cellcolor{15} 100 & \cellcolor{15} 100 & \cellcolor{15} 100 & \cellcolor{15} 100 & \cellcolor{15} 100 & \cellcolor{13} 85 & \cellcolor{15} 100 & \cellcolor{15} 100 & \cellcolor{15} 100 & \cellcolor{15} 100 & \cellcolor{12} 81 & \cellcolor{15} 100 & \cellcolor{15} 100 & \cellcolor{15} 100 & \cellcolor{15} 100 & \cellcolor{15} 100 & \cellcolor{15} 100 & \cellcolor{15} 100 & \cellcolor{15} 100 & \cellcolor{15} 100 & \cellcolor{15} 96 & \cellcolor{15} 100 & \cellcolor{15} 100 & \cellcolor{15} 100 & \cellcolor{15} 100 & \cellcolor{15} 100 \\ 
 & www.gazprombank.ru & \cellcolor{15} 100 & \cellcolor{15} 100 & \cellcolor{15} 100 & \cellcolor{15} 100 & \cellcolor{15} 100 & \cellcolor{15} 100 & \cellcolor{12} 78 & \cellcolor{15} 100 & \cellcolor{15} 100 & \cellcolor{15} 100 & \cellcolor{15} 100 & \cellcolor{15} 100 & \cellcolor{15} 100 & \cellcolor{15} 100 & \cellcolor{15} 100 & \cellcolor{15} 100 & \cellcolor{15} 100 & \cellcolor{15} 100 & \cellcolor{15} 100 & \cellcolor{15} 100 & \cellcolor{15} 100 & \cellcolor{15} 95 & \cellcolor{15} 100 & \cellcolor{15} 99 & \cellcolor{15} 99 & \cellcolor{15} 97 & \cellcolor{15} 99 \\ 
\bottomrule
\end{tabular}}
\end{table*}

Table~\ref{tbl:country-upstream} summarizes the measurement results
between 2022-08-01 and 2023-09-19 per country and domain name. Each cell
shows the share of responses that were not blocked. 

For each country, we select all available probes and query for the A record 
of each domain name using the probe's recursive resolver. To increase 
measurement reliability, we only rely on probes that run on software 
version 3 or higher. Also, we do not show results if we collected responses 
from two \glspl{vp} or less.

Table~\ref{tbl:country-upstream} also shows that there is some form of
\gls{dns} blocking in all countries in the \gls{eu}. At the same time,
however, our measurements show the extent to which blocking occurs
differs widely. For example, while \glspl{vp} in Slovenia experience
frequent blocking for domain names in the first round, no \gls{vp}
experiences blocking for domain names belonging to organizations added
in the later rounds. In comparison, \glspl{vp} in France also experience
blocking for most domain names added later to the sanctions list,
however not as often as the initial list of domain names.  We found
practically no evidence of DNS-based blocking of US-sanctioned Russian
banks or government domains.  Outside of the EU, we find some evidence
of blocking on the small number of media outlets sanctioned by the UK
government.

Interestingly, the block lists and their implementation in member
countries and \glspl{isp} were inconsistent over time.  For example, the
German regulator removed two domain names from their list after their
operators removed sanctioned content\cite{sanktionslisten-fragdenstaat}.
Our measurements show that over-compliance varied as not all \glspl{isp}
stopped blocking the corresponding domain names right away, but after a
few months the sites became reachable by all \glspl{isp} again.

In contrast, the \glspl{isp} in Austria started blocking certain domains
only after a few months, even though they were specified months in
advance and already blocked in Germany. Furthermore, as an example of
under-compliance, the newly registered and sanctioned name for
Sputniknews, \texttt{sputnikglobe.com} has not yet been widely blocked
as of this writing.

Overall, DNS-based blocking is present, but varies from provider to provider.  Domain names that belong to organizations listed in the first
Council of the European Union decision~\cite{cfsp_2022_03_01} are
blocked more often than domain names added later.  45\% of our
\glspl{vp} received at least one blocked response for domain names
related to organizations were listed in the first package (i.e., version
of the sanctions list).  This number decreases with each new round of
packages: from 19\% in the second to 17\% in the fourth round. 

\subsection{Lessons From EduVPN networks}
\label{subsec:resultseduvpn}
Arguably, the users of academic and research networks have a high
expectation, desire, and need for open and unrestricted access to
information.  Internet sanctions however may be at odds with certain
academic pursuits.  Therefore, we also want to evaluate if sanctions
enforcement is present on these networks as well.

We configured 11 measurement \glspl{vm} to connect to each of the
available networks supporting the EduVPN platform.  Each \gls{vm}
tunneled traffic through its connected EduVPN session to a
tunnel gateway using the DNS resolvers provided by the VPN session.  We
validated the DNS responses using Google's public DNS service. The tests
were run on May 09, 2023.

Four of the EduPVN networks fall under the legislation of the \gls{eu}.
These are located in \textit{Germany, Denmark, Finland}, and \textit{the
Netherlands}.  Each of these networks announce a DNS resolver in their
own IP address space.  Our results focus on these four institutions.
The non-European EduVPN sites show little evidence of sanctions
enforcement.

We observed different results in all four European research networks.
The Danish institution exhibited a limited amount of blocking.  All
domain names were resolved as expected. We only observed failed TCP and
HTTP/HTTPS tests on the mirror sites of Sputnik News.

In contrast, a Finish institution exhibited the most complete blocking
with negative TCP and web responses for most news outlets and
corresponding mirror sites. DNS responses however were positive and
valid.

\begin{figure}
  \includegraphics[width=\columnwidth]{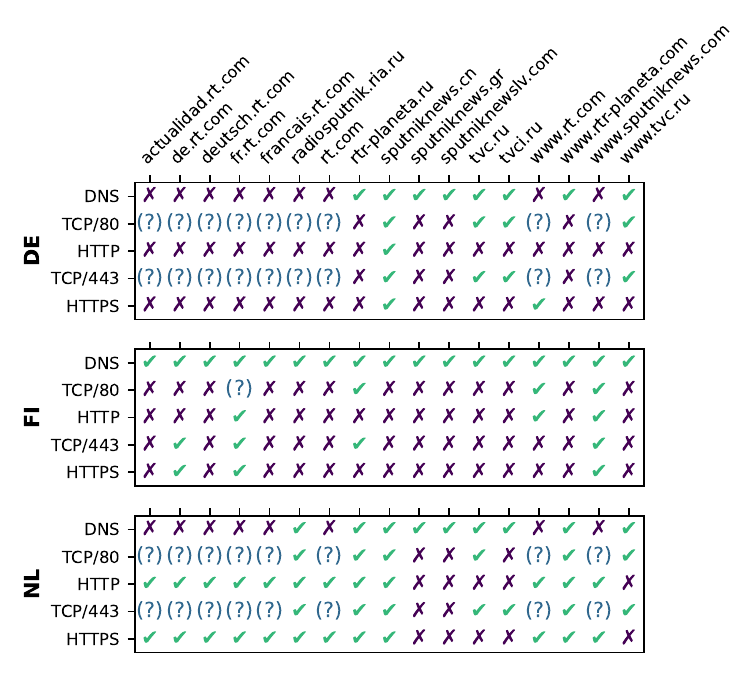}
  \caption{
    Measurement observations of three EduVPN institutions.  Measurements marked with a ``?'' indicate the default limit of 30 hops was exceeded.
  }
  \label{fig:eduvpn-media-outlets}
\end{figure}

The German and a Dutch institutions were similar to one another,
revealing that most media domains were unreachable via TCP and
HTTP/HTTPS.  However, the behavior of the DNS between these networks
differs slightly. In the German network, the resolvers return
\texttt{NXDOMAIN} responses, while the Dutch network answers with
\texttt{SERVFAIL}. We summarize results in Figure
\ref{fig:eduvpn-media-outlets}.

\subsection{Hosting environments and sanctions}
\label{subsec:resultsdaplnlring}
Reachability and HTTP(s) measurements run through Dataplane.org and
NLNOG RING platforms are conducted in a fashion similar to those
performed in the EduVPN environment described above.  We ran HTTP/HTTPS and network connectivity tests on three separate occasions in May 2023. We verified all \glspl{vp} were able to reach at least one of our two control targets. Our tests include retry mechanisms to smooth over any natural, short-lived effects of host, path, and destination variants given the size and diversity of \glspl{vp}.

The results from both Dataplane.org and NLNOG RING are similar, but
NLNOG RING \glspl{vp} were noticeably less reliable and exhibited
greater inconsistency.  Both platforms show high high levels of blocking
to \texttt{sputniknewstv.com} throughout the \gls{eu} region.  However,
while Dataplane.org \glspl{vp} exhibited no serious problems accessing
HTTPS at our control nodes, a number of NLNOG RING \glspl{vp} would
occasionally fail.  We believe a larger proportion of the Dataplane.org
\glspl{vp} fared better due to comparatively smaller average load,
greater available resources, and better than average environmental
stability.

The use of Google DNS on the Dataplane.org platform and local resolver
on NLNOG RING result in relatively few instances of DNS-based sanctions
enforcement.  Therefore, we focus on reachability and HTTP(S) connection
tests.

\begin{figure*}
  \centering
  \includegraphics[width=\textwidth]{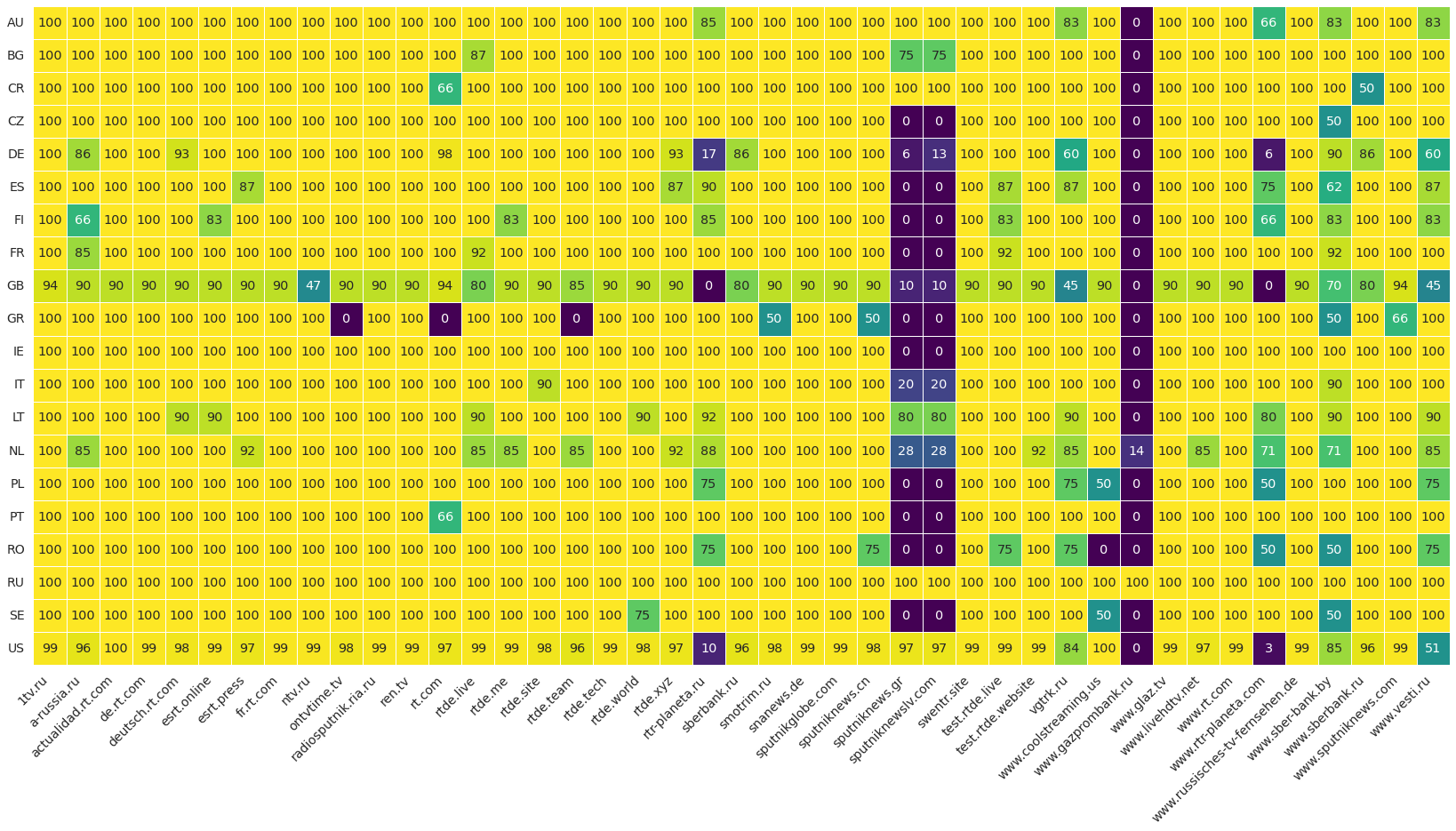}
  \caption{
    HTTPS reachability success rate (2xx or 3xx HTTP response status code) from Dataplane.org VPs.
  }
  \label{fig:dapl_https}
\end{figure*}

Figure \ref{fig:dapl_https} summarizes the success rate of HTTPS
reachability to our sanction list from the Dataplane.org vantage points.
Overall blocking is relatively modest, largely due to the use of Google
DNS, but we find interesting anomalies when we scan the table
vertically.  For example, at the time of measurements,
\texttt{sputniknews.gr} and \texttt{sputniknewslv.com} were both hosted
by a popular DDoS mitigation provider used by many Russian networks.
These domains were largely inaccessible from most of the \gls{eu}.  We
manually verified that traffic between many countries and those sites is
being blocked.  A similar situation appears to have occurred with
\texttt{www.gazprombank.ru}, which was listed in the \gls{usdot}
\gls{ofac} sanctions list.  We don't know the motivation, but these DDoS
mitigation providers appear to have been performing sanctions
enforcement based on IP access control from countries that imposed
sanctions.  It is also worth nothing that these blocks do not show up in
our DNS-based measurements.

\section{Discussion}

\subsection{Transparency of blocking}\label{ssec:transparency}

When analysing our measurements, we noticed network providers convey
vastly different messages to their users when access to a sanctioned
internet resource has been blocked, if they communicate a reason at all,

Overall, the vast majority of \glspl{isp} choose to implement blocking
by some form of DNS-based filtering. RIPE Atlas measurements
\S\ref{subsec:resultsripeatlas} suggest 50\% of \glspl{isp} return a DNS
error response to queries requesting a blocked domain name and in three
\glspl{vp} feedback is conveyed via Extended DNS Errors~\cite{rfc8914}
(info code 15 - blocked).  OONI measurements \S\ref{subsec:resultsooni}
show that 87\% of the 125 \glspl{vp} implementing blocks chose to do so
via DNS. Of these, only 32 serve a block page, meaning that in 74\% of
cases where blocking is implemented the user is not informed of the
reason why the resource is inaccessible.

The usefulness of block pages also varies. Some inform users the domain
name is blocked because of Russia sanctions (e.g.\
Figure~\ref{fig:block_page_sanctions}). Others only show a generic block
page that  are also used for sites blocked due to copyright infringement
(e.g.\ Figure~\ref{fig:block_page_generic}).

\subsection{Mirror pages}

In response to DNS-based sanctions, new Russian domain names were
registered that mirrored the German and Spanish sanctioned websites of
\gls{rt} (see Table~\ref{tbl:domainnames}). In Germany, those domain
names were listed on correspondence by the regulator with groups
representing the local internet
industry~\cite{sanktionslisten-fragdenstaat}.  Additionally, the
Austrian provider Liwest published the Spanish domain names on their
block page~\cite{liwest-blocklist}.

Table~\ref{tbl:country-upstream} shows that in the majority of countries
these new mirror pages are not always blocked. For example, the Spanish
sites are only blocked by Austrian providers, but not in Spain as we
would have expected. On the other hand, German mirror pages are blocked
by most providers in Austria and in Germany with some exceptions.
Measurements for \glspl{vp} in one Portuguese network indicate that some
of the mirrored domain names are blocked only part of the time. A
possible explanation could be a load-balancer that forwards queries to
resolvers with diverging block lists but we could not confirm this
theory.

We found another exception with the domain name \texttt{rtde.live}. This
domain is blocked in Austria but not in Germany whereas the third level
domain names \texttt{test.rtde.live} is only blocked in Germany. This is
in accordance with the sanctions by the German regulator, which  lists
\texttt{test.rtde.live} but not its second level variant.

The list provided by the German regulator also contains domain names not
directly related to the sanctioned organizations, but which facilitate
the distribution of their content. These include websites that allow
visitors to stream the channels of \gls{rt} among others.  These names
change with some regularity and we can use these changes to observe
correlated sanctions enforcement changes.  For example, we saw that
after the German regulator removed the domain names
\texttt{www.russisches-tv-fernsehen.de} and
\texttt{www.coolstreaming.us} from the block list German \glspl{isp}
followed suite accordingly. Our measurements from August 2022 show that
German networks that originally blocked these domain names have lifted
the blocks again.

\subsection{Implementation of sanctions in NRENs}

As already mentioned in section \ref{subsec:resultseduvpn}, researchers
and academics have a keen desire to access otherwise restricted information.
Therefore, we wondered if research facilities would be
excluded from the regulations.  Our work provides evidence that
some \glspl{nren} adhere to sanctions enforcement. The measurements of
a Finnish \gls{nren} reveal a rather broad implementation of sanctions.
Our observation of the German and Dutch networks align with DNS-based
enforcement of regular German ISPs.  Only a research network in Denmark
appears to be less strict in comparison to national ISPs.
A comparison of Table \ref{tbl:country-upstream} and Figure
\ref{fig:eduvpn-media-outlets} support these findings.

\subsection{Placement of enforcement mechanisms}

We found DNS-based blocking was the dominant form of implementing
internet sanctions.  Functionally however, networks vary widely in how
DNS-based blocking was performed.  Some networks redirect users to a
page that may or may not explain why a resource was blocked.  Others
simply return DNS errors, sometimes with extended error codes, but
usually not.  Many networks rely on third party resolution service such
as Google Public DNS or Cloudflare DNS, which do not appear to implement
any sanctions enforcement regardless of location.  As long as a user
can utilize an alternative DNS resolver, they would be able to bypass
most sanctions enforcement.

We found some evidence of IP address access control to enforce sanctions.
These mechanisms were typically the most complete and successful because
they were applied close to or at the destination where the restricted content
was hosted.  While this approach was most effective, it was also the the
least popular type of mechanism deployed.  This approach would also pose
the greatest risk of "over-blocking" when multiple systems and services
share an IP address, which may explain why it was rarely employed.

\section{Related Work}

The Russian invasion in the Ukraine and immediate consequences on
internet connectivity has sparked the interest of the research
community.  In addition to generally network availability issues,
internet censorship was studied in the context of the war of Ukraine.
However, to the best of our knowledge, no studies focus on internet
sanctions within the \gls{eu}. The OONI project team has measured
censorship within Russia and show that censorship was extended to a
broader set of sites and services in the course of the
conflict~\cite{xynou2022, roskomsvoboda2023}. Also Ramesh et al.\ show
that censorship on Russian users increased.~\cite{ramesh2022}.
Additionally, they study the use of a new domestic certificate authority
and the use and blocking of censorship circumvention tools.   

Other literature and reports focused on the response of services and
infrastructure in the Ukraine and in Russia. Jonker et al.\
\cite{jonker2022ru} study how the infrastructure of Russian websites and
\gls{dns} infrastructure changed in the course of the conflict. Luconi
et al.\ study the impact on routing and latency in the
Ukraine~\cite{luconi2023}.

Outside of the scope of the Russian invasion, online censorship has been
studied extensively. Poort et al.\ study censorship and impact in
Europe in the context of copyright
infringement~\cite{poort2014baywatch}. Their focus was on \glspl{isp} in
the Netherlands and not on the technical implementation of the measures.
Bortzmeyer has used RIPE Atlas probes to measure censorship
worldwide~\cite{bortzmeyer2015}. There, he also discusses caveats when
using RIPE Atlas probes to measure censorship. Ververis et al.\ study
the impact of censorship in mobile app
stores~\cite{ververis_shedding_2019}.  The work of Ververis et al.\ \cite{ververis2023website} comes close to our own. Using OONI they
consider censorship generally, examining block list consistency and web censorship behavior throughout EU member countries.

Finally, the topic of digital sovereignty has also gained traction in
the social science community. Perarnaud et al.\ analyse various EU
policies on digital sovereignty and their
impacts~\cite{perarnaud_splinternets_2022}\mmu{Nils, could you highlight
how this work differs to ours and check that the rest of the paragraph
is alright?}. Braud et al.\ and Baur et al.\ both analysed digital
sovereignty by looking at the ``European Cloud'' project
Gaia-X~\cite{braud_road_2021,baur_european_2023}. With our analysis we
add a novel perspective on the issue of digital sovereignty in the
\gls{eu} and for the future of internet sanctions.

\section{Conclusions}

In this article we have analyzed how EU sanctions against Russian media,
in response to the Russian aggression in Ukraine, have been implemented.
What we have found is that these sanctions are inconsistently
implemented across the EU.  The inconsistent implementation of the
sanctions can at least in part be attributed to the high-level
description of the sanctions and the lack of recommendations for
technical implementation. This left implementation of the sanctions
largely to the interpretation of network operators without sufficient
guidance provided by national authorities in EU member states.  This led
to a diffuse implementation and thus it could be argued has had limited
impact to increasing the EU's digital sovereignty. However, it could be
also be said that this is typical for EU policy making, which always
involves raised tensions when it comes to the sovereignty of individual
member states. 

While the sanctions might not have proven to be as efficient and
effective as some may have liked, we anticipate this is just an
early harbinger into a new era of multilateral internet sanction events.
This may also accelerate the discursive concept of digital sovereignty
into a technical reality. Future work could compare the implementation
of sanctions with other approaches to attain digital sovereignty,
specifically in the EU due to its supranational nature and inherent
tension between governance layers. 

\label{lastpage}


\bibliographystyle{ACM-Reference-Format}
\bibliography{library,rfc}

\appendix

\section{Block pages, timeline, sanctions list}
\label{appendix:blockpages}

\begin{figure}[!ht]
  \centering
  \includegraphics[width=\columnwidth]{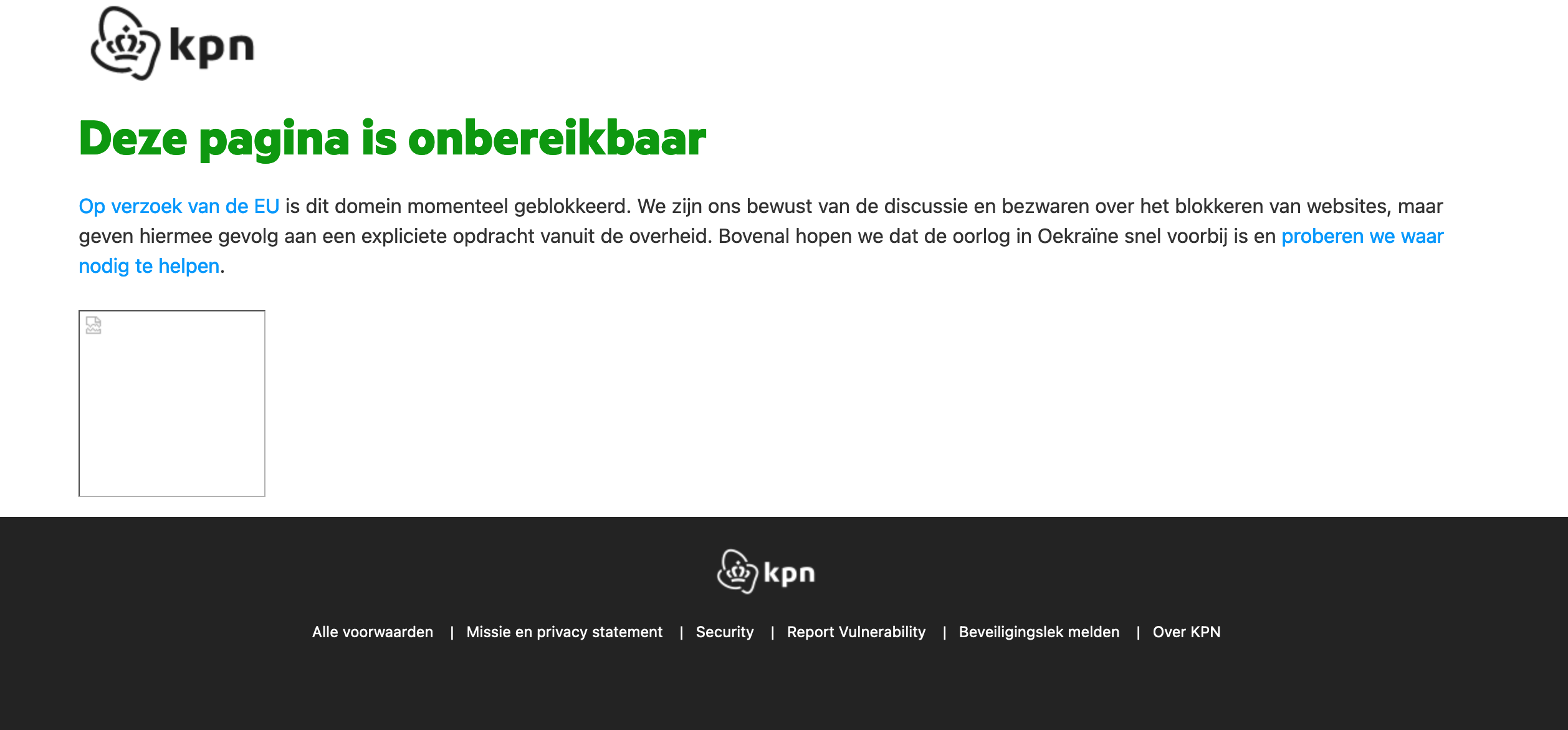}
  \caption{Blocking page of a Dutch ISP. The text states ``Upon the request of
  the EU [link to the Dutch version of~\cite{cfsp_2022_03_01}], this domain is
  currently blocked. We are aware of the discussion and objections about
  blocking websites, but we are complying with an explicit order from the
  government. Above all, we hope that the war in Ukraine will soon be over and
  we will try to help where needed.'' (translated by the
  authors).}\label{fig:block_page_sanctions}
\end{figure}

\begin{figure}[H]
  \centering
  \includegraphics[width=0.8\columnwidth]{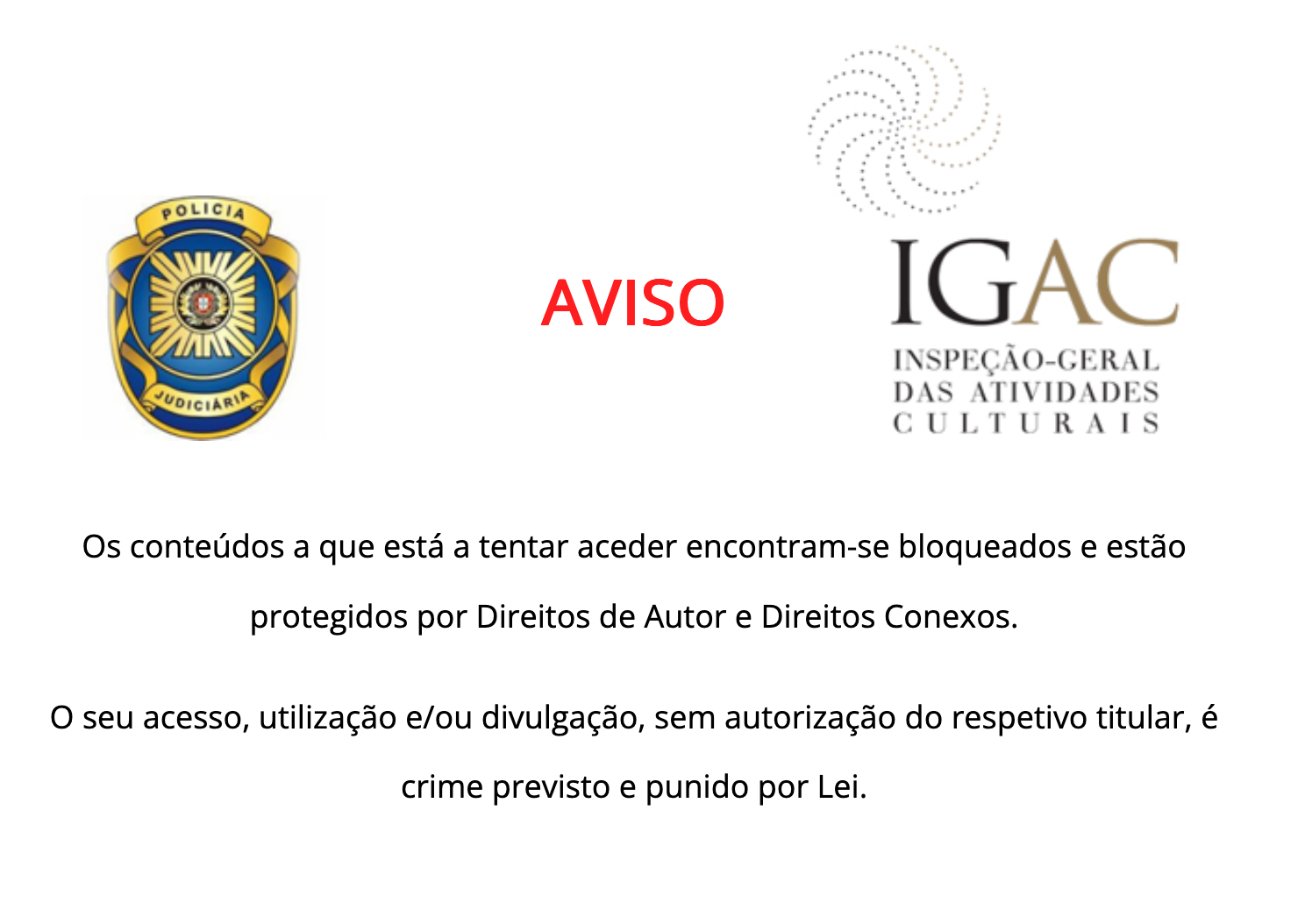}
  \caption{Blocking page of a Portuguese ISP. The text states ``The contents you are trying to access are blocked and are protected by Copyright and Related Rights. Its access, use and/or disclosure, without the authorization of the respective holder, is a crime provided for and punished by law.'' (translated by Google Translate).}\label{fig:block_page_generic}
\end{figure}

\begin{figure}[H]
  \centering
  \includegraphics[width=0.8\columnwidth]{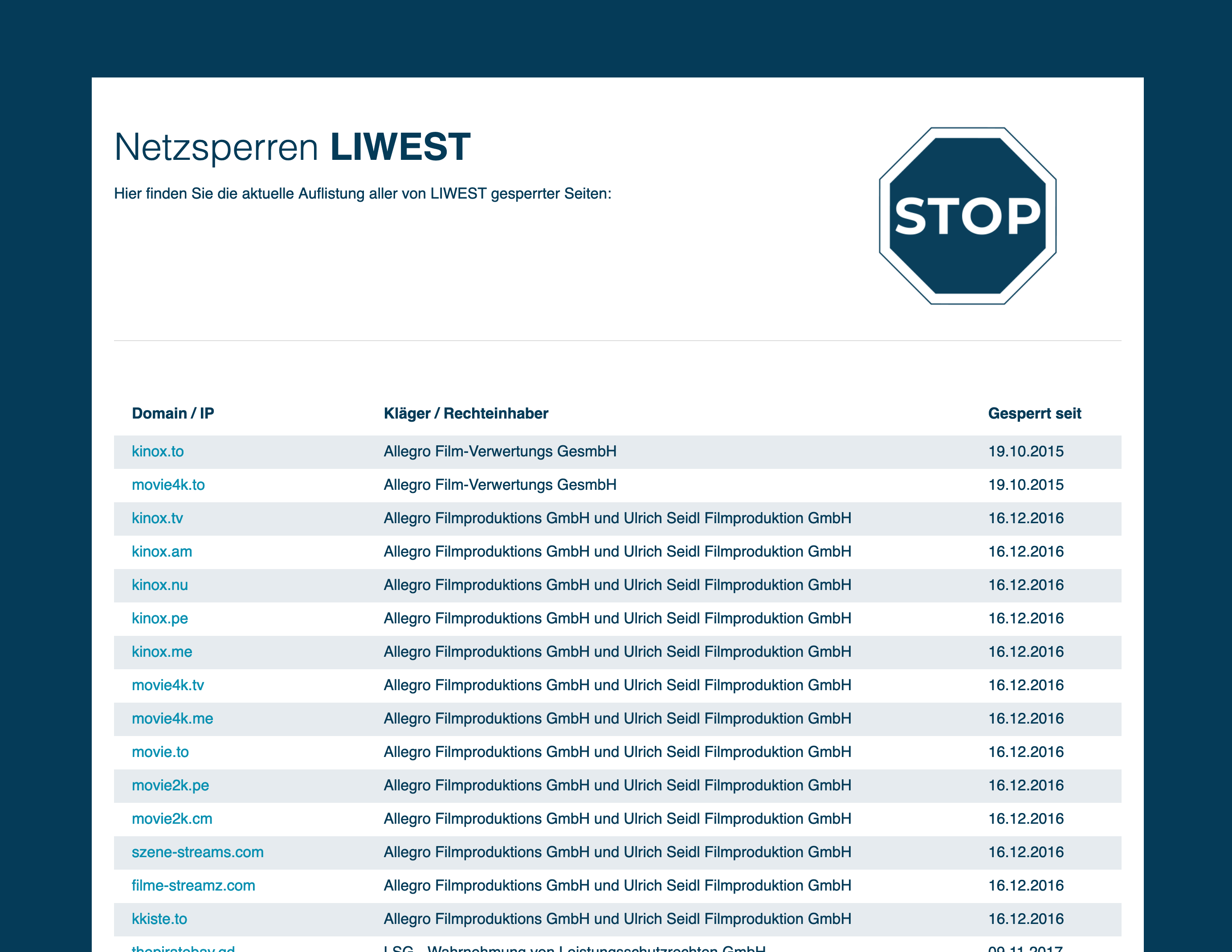}
  \caption{Blocking page of an Austrian ISP. The page lists all currently blocked pages, the entity requesting the block, and the date the block was added.}\label{fig:block_page_sanctions_detailed}
\end{figure}

\clearpage
\label{appendix:sanctioned_domain_history}
\begin{table*}
    \caption{Sanctioned organisation, measured hostnames, and source.}\label{tbl:domainnames}
    \centering
    \adjustbox{max width=\textwidth}{%
    \begin{tabular}{l l l p{5cm}}
    \toprule
    Sanctioned organisation                                         & Hostname                              & Source                                                & Remark/Date added   \\ \midrule       
    Russia Today English                                            & www.rt.com                            & Council Decision 2022/351~\cite{cfsp_2022_03_01}      & 1 March 2022               \\
    Russia Today UK                                                 & www.rt.com                            & Council Decision 2022/351~\cite{cfsp_2022_03_01}      & 1 March 2022               \\
    Russia Today Germany                                            & de.rt.com                             & Council Decision 2022/351~\cite{cfsp_2022_03_01}      & 1 March 2022               \\
                                                                    & deutsch.rt.com                        & Council Decision 2022/351~\cite{cfsp_2022_03_01}      & 1 March 2022               \\
    Russia Today France                                             & francais.rt.com                       & Council Decision 2022/351~\cite{cfsp_2022_03_01}      & 1 March 2022               \\
                                                                    & fr.rt.com                             & Council Decision 2022/351~\cite{cfsp_2022_03_01}      & 1 March 2022               \\
    RT en español                                                   & actualidad.rt.com                     & Council Decision 2022/351~\cite{cfsp_2022_03_01}      & 1 March 2022               \\
                                                                    & actualidad-rt.com                     & Council Decision 2022/351~\cite{cfsp_2022_03_01}      & 1 March 2022    \\ 
    Sputnik                                                         & www.sputniknews.com                   & Council Decision 2022/351~\cite{cfsp_2022_03_01}      & 1 March 2022               \\
                                                                    & sputniknewslv.com                     & Council Decision 2022/351~\cite{cfsp_2022_03_01}      & 1 March 2022               \\
                                                                    & sputniknews.gr                        & Council Decision 2022/351~\cite{cfsp_2022_03_01}      & 1 March 2022               \\
                                                                    & sputniknews.cn                        & Council Decision 2022/351~\cite{cfsp_2022_03_01}      & 1 March 2022               \\
                                                                    & radiosputnik.ria.ru                   & Council Decision 2022/351~\cite{cfsp_2022_03_01}      & 1 March 2022               \\
                                                                    & sputnikglobe.com                      & Council Decision 2022/351~\cite{cfsp_2022_03_01}      & Registered 29 March 2023, sputniknews.com now redirects to this domain name.  \\
    Rossiya RTR / RTR Planeta                                       & www.rtr-planeta.com                   & Council Decision 2022/884~\cite{cfsp_2022_06_03}      & 3 June 2022 \\
                                                                    & rtr-planeta.ru                        & Council Decision 2022/884~\cite{cfsp_2022_06_03}      & 3 June 2022 \\
                                                                    & vgtrk.ru                              & Council Decision 2022/884~\cite{cfsp_2022_06_03}      & 3 June 2022 \\
    Rossiya 24 / Russia 24                                          & www.vesti.ru                          & Council Decision 2022/884~\cite{cfsp_2022_06_03}      & 3 June 2022 \\
    TV Centre International                                         & www.tvc.ru                            & Council Decision 2022/884~\cite{cfsp_2022_06_03}      & 3 June 2022 \\
                                                                    & tvci.ru                               & Council Decision 2022/884~\cite{cfsp_2022_06_03}      & 3 June 2022 \\
    NTV/NTV Mir                                                     & ntv.ru                                & Council Decision 2022/2478~\cite{cfsp_2022_12_16}     & 16 December 2022 \\
    Rossiya 1                                                       & smotrim.ru                            & Council Decision 2022/2478~\cite{cfsp_2022_12_16}     & 16 December 2022 \\
    REN TV                                                          & ren.tv                                & Council Decision 2022/2478~\cite{cfsp_2022_12_16}     & 16 December 2022 \\
    Pervyi Kanal                                                    & 1tv.ru                                & Council Decision 2022/2478~\cite{cfsp_2022_12_16}     & 16 December 2022 \\
    RT Arabic                                                       & www.rtarabic.com                      & Council Decision 2023/434~\cite{cfsp_2023_02_25}      & 25 February 2023  \\
    Sputnik Arabic                                                  & sputnikarabic.ae                      & Council Decision 2023/434~\cite{cfsp_2023_02_25}      & 25 February 2023 \\ \midrule
    RT en español mirror                                            & esrt.online                           & Liwest Blocklist~\cite{liwest-blocklist}              & Registered 8 April 2022  \\
                                                                    & esrt.press                            & Liwest Blocklist~\cite{liwest-blocklist}              & Registered 8 April 2022   \\
    RT Germany mirror                                               & rtde.site                             & Bundesnetzagentur~\cite{sanktionslisten-fragdenstaat} & Registered 5 March 2022 \\
                                                                    & rtde.xyz                              & Bundesnetzagentur~\cite{sanktionslisten-fragdenstaat} & Registered 5 March 2022  \\
                                                                    & rtde.team                             & Bundesnetzagentur~\cite{sanktionslisten-fragdenstaat} & Registered 5 March 2022  \\
                                                                    & test.rtde.live                        & Bundesnetzagentur~\cite{sanktionslisten-fragdenstaat} & Registered 6 April 2022  \\
                                                                    & rtde.live                             & Bundesnetzagentur~\cite{sanktionslisten-fragdenstaat} & Registered 6 April 2022  \\
                                                                    & test.rtde.website                     & Bundesnetzagentur~\cite{sanktionslisten-fragdenstaat} & Registered 6 April 2022 \\
                                                                    & rtde.tech                             & Liwest Blocklist~\cite{liwest-blocklist}              & Registered 6 April 2022   \\
                                                                    & rtde.world                            & Liwest Blocklist~\cite{liwest-blocklist}              & Registered 6 April 2022  \\
                                                                    & rtde.me                               & Liwest Blocklist~\cite{liwest-blocklist}              & Registered 6 April 2022  \\  \midrule
    A-Russia                                                        & a-russia.ru                           & Bundesnetzagentur~\cite{sanktionslisten-fragdenstaat} & Russian TV streaming site \\  
    WWITV: World Wide Internet TV                                   & wwitv.com                             & Bundesnetzagentur~\cite{sanktionslisten-fragdenstaat} & TV streaming site \\  
    glaz.tv                                                         & www.glaz.tv                           & Bundesnetzagentur~\cite{sanktionslisten-fragdenstaat} & TV streaming site \\  
    Russisches Fernsehen                                            & www.russisches-tv-fernsehen.de        & Bundesnetzagentur~\cite{sanktionslisten-fragdenstaat} & TV streaming site \\  
    On TV Time                                                      & ontvtime.tv                           & Bundesnetzagentur~\cite{sanktionslisten-fragdenstaat} & TV streaming site \\  
    SPB TV World                                                    & spbtv.online                          & Bundesnetzagentur~\cite{sanktionslisten-fragdenstaat} & TV streaming site \\  
    Coolstreaming                                                   & www.coolstreaming.us                  & Bundesnetzagentur~\cite{sanktionslisten-fragdenstaat} & TV streaming site \\  
    Live HD TV                                                      & www.livehdtv.net                      & Bundesnetzagentur~\cite{sanktionslisten-fragdenstaat} & TV streaming site \\  
    Rossiya Segodnya Group                                          & snanews.de                            & Liwest Blocklist~\cite{liwest-blocklist}              & German news site \\ \midrule
    State Duma                                                      & duma.gov.ru                           & OFAC Sanctions list~\cite{ofac_2023}                        &                   \\
    Sberbank                                                        & www.sber-bank.by                      & Council Decision 2022/327~\cite{cfsp_2022_02_25}      & 25 February 2022, Not part of Annex IX \\
                                                                    & www.sberbank.ru                       & Council Decision 2022/327~\cite{cfsp_2022_02_25}      & 25 February 2022, Not part of Annex IX \\
    Gazprombank                                                     & www.gazprombank.ru                    & Council Decision 2022/2478~\cite{cfsp_2022_12_16}     & 16 December 2022, Not part of Annex IX \\
    \bottomrule
    \end{tabular}}
\end{table*}



\label{appendix:sanctions_action_time}
\begin{figure}
  \centering
  \includegraphics[width=0.8\columnwidth]{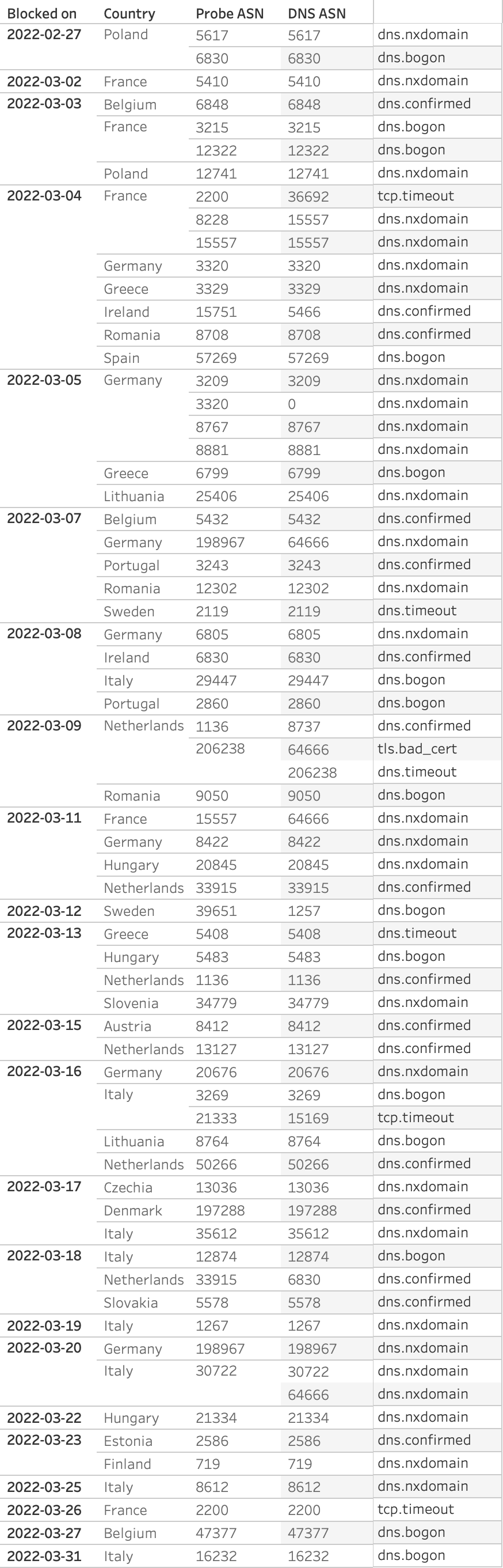}
  \caption{Timeline of blocking of www.rt.com across providers in Europe}\label{fig:ooni_timeline_blocks1}
\end{figure}

\begin{figure}
  \centering
  \includegraphics[width=0.8\columnwidth]{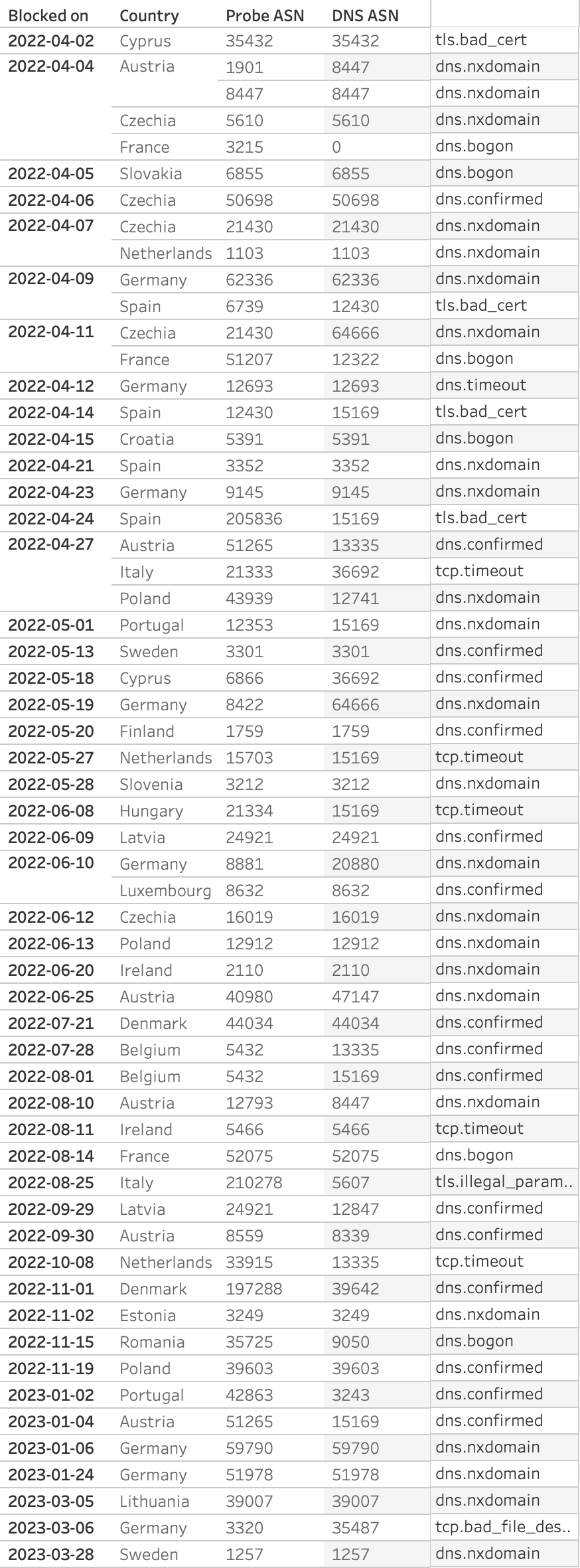}
  \caption{Timeline of blocking of www.rt.com across providers in Europe}\label{fig:ooni_timeline_blocks2}
\end{figure}

\end{document}